\documentclass[11pt,a4paper]{article}
\pdfoutput=1
\usepackage{mysty}
\usepackage{amsfonts}
\usepackage{amsmath}
\usepackage{cancel}
\usepackage{graphicx}
\usepackage{hyperref}
\usepackage{multirow}
\usepackage{makecell}
\usepackage[table]{xcolor}
\usepackage{xspace}
\usepackage{placeins}
\usepackage{lmodern}
\usepackage[left=1.85in, right=-0.2in,top=1.8in,bottom=-0.2in]{geometry}

\definecolor{colorDG}{HTML}{008000}

\newcommand{\tp}[1]{{\color{red}  #1}}

\newcommand{\RF}[1]{{\color{blue}  #1}}

\newcommand{\nl}{{\mathnormal l}}
\newcommand{\rn}{{\tt r}}
\newcommand{\Mn}{{\tt M}}
\newcommand{\R}{\text{Re}}
\newcommand{\I}{\text{Im}}

\newcommand{\wi}{\text{with}}
\newcolumntype{P}[1]{>{\centering\arraybackslash}p{#1}}

\hypersetup{pdfstartview=FitV,colorlinks=true,linkcolor=blue,citecolor=red,filecolor=black,urlcolor=blue}

\definecolor{color1}{HTML}{00dd00}
\definecolor{color2}{HTML}{dd0000}

\definecolor{color8TeV}{HTML}{ddddff}
\definecolor{color13TeV}{HTML}{eeee99}
\definecolor{color2TeV}{HTML}{9DCEBF}

\allowdisplaybreaks

\begin{document}

\title{Updated global fit of the aligned two-Higgs-doublet model with heavy scalars}

\author[a]{Anirban Karan,}
\emailAdd{kanirban@ific.uv.es}

\author[b]{V\'ictor Miralles,}
\emailAdd{victor.miralles@roma1.infn.it}

\author[a]{Antonio Pich}
\emailAdd{antonio.pich@ific.uv.es}

\affiliation[a]{Institut de F\'isica Corpuscular (CSIC-UV), Parque Cient\'ifico, Catedr\'atico Jos\'e Beltr\'an 2, E-46980 Paterna, Spain}

\affiliation[b]{INFN, Sezione di Roma, Piazzale A. Moro 2, I-00185 Roma, Italy}

\abstract{An updated global fit on the parameter-space of the Aligned Two-Higgs-Doublet model has been performed with the help of the open-source package {\tt HEPfit}, assuming the Standard-Model Higgs to be the lightest scalar. No new sources of CP violation, other than the phase in the CKM matrix of the Standard Model, have been considered. A similar global fit was previously performed in Ref. \cite{Eberhardt:2020dat} with a slightly different set of parameters. Our updated fit incorporates improved analyses of the theoretical constraints required for perturbative unitarity and boundedness of the scalar potential from below, additional flavour observables and updated data on direct searches of heavy scalars at the LHC, Higgs signal strengths and electroweak precision observables. Although not included in the main fit, the implications of the CDF measurement of the $W^\pm$ mass are also discussed.

}

\preprint{{\raggedleft IFIC/23-30 \par}}

\maketitle
\vspace*{0mm}
\section{Introduction}
\label{sec:Intro}

With the discovery of the Higgs boson \cite{ATLAS:2012yve,CMS:2012qbp}, the Standard Model (SM) has become a well-established theory describing the interactions among elementary particles in a very elegant way. Nonetheless, there are ample evidences indicating that the SM cannot explain all aspects of nature, and hence the existence of physics beyond the Standard Model (BSM) is indispensable. 
 In order to explain such BSM phenomena, the SM is usually enlarged 
 with some additional particles or gauge groups that respect all its fundamental principles. 
 While the augmentation of the SM with extra generations of quarks and leptons receives severe experimental constraints from unitarity-triangle data \cite{UTfit:2022hsi} or $Z$-boson branching fractions \cite{ALEPH:2005ab}, extensions with additional $SU(2)_L$ scalar doublets do not suffer from such stringent bounds.\footnote{In generic scalar extensions the tree-level value of the $\rho$ parameter shifts from unity, depending on the hypercharge and weak isospin of the extra scalars \cite{Diaz-Cruz:2003kcx}. However, it remains equal to 1
with additional doublets having the same hypercharge as the SM Higgs doublet.}
	
Among such scalar extensions, the simplest one is the two-Higgs-doublet model (THDM),  where one additional scalar doublet with the same quantum numbers than the SM Higgs doublet is appended \cite{Branco:2011iw,Gunion:1989we,Ivanov:2017dad}. In addition to the three needed Goldstone bosons, this model includes three neutral and one pair of charged scalars. Such rich scalar spectrum opens
various interesting possibilities such as new sources of CP violation \cite{Gunion:2005ja, Wu:1994ja, Keus:2015hva,Chen:2017com,Iguro:2019zlc}, axion-like phenomenology \cite{Kim:1986ax, Espriu:2015mfa, Celis:2014zaa}, dark matter aspects \cite{LopezHonorez:2006gr,Belyaev:2016lok,Tsai:2019eqi}, neutrino mass generation \cite{Ma:2006km, Hirsch:2013ola}, electroweak baryogenesis \cite{Turok:1990zg, Cline:2011mm,Fuyuto:2015jha}, stability of the scalar potential till the Planck scale \cite{Ferreira:2015rha,Das:2015mwa,Schuh:2018hig}, etc. Moreover, the THDM can also be thought of as a low-energy effective field theory framework for several models with larger symmetry groups (like supersymmetry).
	
One major shortcoming of the most general THDM is the emergence of tree-level flavour-changing neutral currents (FCNC), which are observed to be tightly constrained experimentally.
This problem is usually avoided by imposing
discrete $\mathcal Z_2$ symmetries 
so that each type of right-handed fermions couples to one scalar doublet only \cite{Glashow:1976nt,Paschos:1976ay}.
However, the absence of tree-level FCNC can be guaranteed 
with a much weaker requirement: the alignment of Yukawa couplings in the flavour space, so that the interactions of the two scalar doublets acquire the same flavour structure \cite{Pich:2009sp,Pich:2010ic,Penuelas:2017ikk,Manohar:2006ga}. This provides a more generic theoretical framework, known as  
the `Aligned Two-Higgs Doublet model' (ATHDM), where flavour violation is minimal \cite{Chivukula:1987py,DAmbrosio:2002vsn} and (highly-suppressed) FCNC only appear at higher perturbative orders \cite{Pich:2009sp,Pich:2010ic,Ferreira:2010xe,Jung:2010ik,Braeuninger:2010td,Bijnens:2011gd,Li:2014fea,Abbas:2015cua,Botella:2015yfa,Penuelas:2017ikk,Gori:2017qwg}. In this model, all the fermion-scalar interactions become proportional to the masses of the corresponding fermions, leading to a
quite compelling 
phenomenology in  high-energy colliders \cite{ Abbas:2015cua, Celis:2013ixa, Celis:2013rcs} as well as in low-energy flavour experiments \cite{Jung:2010ik,Jung:2010ab,Jung:2012vu}. The ATHDM constitutes a generic platform for THDMs; all $\mathcal{Z}_2$-symmetric THDM scenarios can be explored as special cases of the ATHDM \cite{Pich:2009sp}. In addition, the ATHDM can accommodate new sources of CP violation beyond the CKM matrix in both the scalar and Yukawa sectors \cite{Pich:2009sp}.
	
The parameter space of the THDM has been extensively scrutinized in the literature, considering LHC data, LEP data, flavour bounds and theoretical constraints \cite{Eberhardt:2020dat,Jung:2010ik,Celis:2013rcs,Abbas:2015cua,Celis:2013ixa,Ilisie:2014hea,Chowdhury:2017aav, Haller:2018nnx, Cacchio:2016qyh,Chowdhury:2015yja,Eberhardt:2013uba,Wang:2013sha,Botella:2015hoa,Craig:2015jba,Bernon:2014nxa,Bernon:2015qea,Ilnicka:2015jba,Bernon:2015wef,Belusca-Maito:2016dqe,Dercks:2018wch,Ilnicka:2018def,Sanyal:2019xcp,Herrero-Garcia:2019mcy,Karmakar:2019vnq,Chen:2019pkq,Arco:2020ucn,Botella:2018gzy,Botella:2020xzf,Botella:2022rte,Botella:2023tiw,Athron:2021auq, Connell:2023jqq, Karan:2023xze}. A global fit of the ATHDM with no new sources of CP violation (and a slightly different set of parameters) was performed in Ref.~\cite{Eberhardt:2020dat}, taking into account various theoretical and experimental bounds. In this paper we reanalyse the global fit of this scenario in great detail, assuming that the SM Higgs boson is the lightest scalar. The fits are performed with the help of the open-source code {\tt HEPfit}  \cite{DeBlas:2019ehy}. Compared with the previous study, we have updated the Higgs signal strengths and added direct searches for all the scalars from the most recent LHC data. Additionally, we have improved the analysis of theoretical constraints by including \textit{necessary and sufficient} conditions for boundedness of the scalar potential from below (i.e. the potential never tends to negative infinity), and incorporated the branching fractions for semileptonic decays of $B$, $D$ and $K$ mesons in the flavour sector too. We have also analysed the 
implications of the CDF measurement of the $W^\pm$ mass \cite{CDF:2022hxs}, 
although we have not included it in our global fits in view of its currently unresolved discrepancy with other measurements. A similar conservative attitude has been taken for
the muon $g-2$ anomaly \cite{Aoyama:2020ynm}, in view of the current controversy with lattice \cite{Borsanyi:2020mff,Ce:2022kxy,ExtendedTwistedMass:2022jpw} and $\tau$-decay \cite{Davier:2010fmf,Pich:2013lsa} data; although not included in the global fit, we have also studied the ensuing constraints on the parameter space of the ATHDM.
 
The paper is organized in the following way. Next section (Section \ref{sec:Model}) briefly describes the ATHDM scenario. The set up for the global fit is discussed in Section \ref{sec:set_up}. While in Section \ref{sec:constraints}, we illustrate different theoretical as well as experimental constraints that are considered for this study, in the subsequent section (Section \ref{sec:results}) we present the numerical results from the fits. Finally we conclude in Section \ref{sec:conclusion}. All the data and various other details used in the global fit are incorporated in the Appendix section.

\section{The ATHDM model}

\label{sec:Model}

\subsection{Scalar Sector}

We extend the SM with a second complex scalar doublet having the same hypercharge as the SM scalar doublet, i.e. $Y=1/2$. 
The neutral components of both scalar doublets may acquire non-zero complex vacuum expectation values (vev); nonetheless, with a suitable $SU(2)_L\otimes U(1)_Y$ global transformation, one can always rotate the basis of the scalar space such that only the neutral component of the first doublet acquires a non-zero (real) vev.
Working in this so-called Higgs-basis, we can write down the two doublet scalar fields 
as:
\begin{equation}
\Phi_1=\frac{1}{\sqrt 2}\begin{pmatrix}
\sqrt 2\;G^+\\
S_1+v+i\, G^0
\end{pmatrix}\, ,\qquad\qquad \Phi_2=\frac{1}{\sqrt 2}\begin{pmatrix}
\sqrt 2\;H^+\\
S_2+i\, S_3
\end{pmatrix}\, ,
\end{equation} 
where $\Phi_1$ gets the vev  $v=246$ GeV. The components $G^\pm$ and $G^0$ act as Goldstone bosons, providing the masses to the $W^\pm$ and $Z$ bosons. Thus, we are left with one pair of charged scalars $H^\pm$, two neutral scalars $S_{1,2}$ and one neutral pseudoscalar $S_3$. The three neutral particles $S_j$ mix with each other through an orthogonal transformation to produce the mass eigenstates $\varphi_i^0$.
The explicit form of this orthogonal transformation depends on the scalar potential. 
In general, if the scalar potential is not invariant under the CP symmetry,
all the three $S_j$ fields mix together giving no definite CP to the mass eigenstates. 

The most general scalar potential, allowed by the SM gauge symmetry, takes the form:
\begin{align}
\label{eq:pot}
V&=\mu_1\,\Phi_1^\dagger \Phi_1+\mu_2\,\Phi_2^\dagger \Phi_2+ \Big[\mu_3\,\Phi_1^\dagger \Phi_2+h.c.\Big]+\frac{\lambda_1}{2}\,(\Phi_1^\dagger \Phi_1)^2+\frac{\lambda_2}{2}\,(\Phi_2^\dagger \Phi_2)^2+\lambda_3\,(\Phi_1^\dagger \Phi_1)(\Phi_2^\dagger \Phi_2)\nonumber\\
&+\lambda_4\,(\Phi_1^\dagger \Phi_2)(\Phi_2^\dagger \Phi_1)+\Big[\Big(\frac{\lambda_5}{2}\,\Phi_1^\dagger \Phi_2+\lambda_6 \,\Phi_1^\dagger \Phi_1 +\lambda_7 \,\Phi_2^\dagger \Phi_2\Big)(\Phi_1^\dagger \Phi_2)+ \mathrm{h.c.}\Big]\, ,
\end{align}
where $\mu_3,\lambda_5,\lambda_6$ and $\lambda_7$ are complex parameters and the rest are real. Minimizing the potential with respect to the neutral fields at their corresponding minima, one obtains:
\begin{equation}
v^2=-\frac{2\mu_1}{\lambda_1}=-\frac{2\mu_3}{\lambda_6}\, ,
\end{equation}
implying that $\mu_1$ and $\mu_3$ are not independent quantities. Redefining the phase of $\Phi_2$, 
one can make any one of the three parameters $\lambda_5,\lambda_6$ or $\lambda_7$ real. Therefore, 
the scalar potential involves eleven independent real
parameters: $\mu_2,\; v,\; \lambda_{1,2,3,4},\; |\lambda_{5,6,7}|$ and the two relative phases between $\lambda_{5,6,7}$. 

The quadratic terms of the potential, 
which give rise to the scalar masses, can be written as:
\begin{equation}
    V_M = \left(\mu_2+\frac{1}{2}\lambda_3 v^2\right) H^+H^-+\frac{1}{2}\begin{pmatrix} S_1 & S_2 & S_3 \end{pmatrix} \mathcal{M} \begin{pmatrix} S_1 \\ S_2 \\ S_3 \end{pmatrix}\, ,
\end{equation}
with   
\begin{equation}
\mathcal{M}=\begin{pmatrix} v^2\, \lambda_1  && v^2\, \rm{Re}(\lambda_6) && -v^2\, \rm{Im}(\lambda_6) \\
    v^2\, \rm{Re}(\lambda_6) && \mu_2+\frac{1}{2}v^2\{\lambda_3+\lambda_4+\rm{Re}(\lambda_5)\} && -\frac{1}{2}v^2\, \rm{Im}(\lambda_5) \\
    -v^2\, \rm{Im}(\lambda_6) && -\frac{1}{2}v^2\, \rm{Im}(\lambda_5) && \mu_2+\frac{1}{2}v^2\{\lambda_3 + \lambda_4 - \rm{Re}(\lambda_5) \} \end{pmatrix}\, .
    \label{eq:massmatrix}
\end{equation}

In order to obtain the physical mass eigenstates,
we need to diagonalise the neutral mass matrix~\eqref{eq:massmatrix}. Imposing CP conservation in the scalar sector, all the parameters in our potential become real, reducing to nine the number of independent inputs.
Moreover, in this case the mass eigenstates have definite CP: we get two CP-even ($h$, $H$) and one CP-odd ($A$) fields.
From the mass matrix, we explicitly see that in this case $S_3=A$ does not mix with the other scalars
and we only need to diagonalise a $2\times 2$ matrix.
The squared masses are then given by:
\begin{equation}
M_{H^\pm}^2=\mu_2+\frac{\lambda_3}{2}\,v^2\, ,\qquad M_{h,H}^2=\frac{1}{2}\,(\Sigma\mp\Delta),\qquad M_A^2=M_{H^\pm}^2+\frac{v^2}{2}\,(\lambda_4-\lambda_5)\, ,
\end{equation}
with
\begin{equation}
\Sigma=M_{H^\pm}^2+\Big(\lambda_1+\frac{\lambda_4}{2}+\frac{\lambda_5}{2}\Big)\,v^2\qquad \text{ and }\qquad \Delta=\sqrt{\big(\Sigma-2\lambda_1 v^2\big)^2+4\,\lambda_6^2\,v^4}\, .
\end{equation}
Note also that, since the trace of the mass matrix must be invariant, the scalar masses must satisfy the following relation:
\begin{equation}
    M_{h}^2+M_{H}^2+M_{A}^2=2 M_{H^\pm}^2+v^2(\lambda_1+\lambda_4)\, ,
    \label{eq:tracematrix}
\end{equation}
which can also be easily confirmed by substituting the expressions of the masses.
The mixing between the two CP-even neutral scalars,
\begin{equation}
\begin{pmatrix}
h\\H
\end{pmatrix}=\begin{pmatrix}
\cos\tilde{\alpha}& \sin\tilde{\alpha}\\ -\sin\tilde{\alpha}&\cos\tilde{\alpha}
\end{pmatrix}\begin{pmatrix}
S_1\\S_2
\end{pmatrix} 
\, ,
\end{equation}
%
is given by:
\begin{equation}
\tan\tilde\alpha=\frac{M_h^2 -v^2\,\lambda_1}{v^2\,\lambda_6}=\frac{v^2\,\lambda_6}{v^2\,\lambda_1-M_H^2}\, .
\label{eq:mix_ang_lam}
\end{equation}

 Using the above equations,  we can trade five of the nine parameters of the potential ($\mu_2$, $v$ and $\lambda_{1-7}$) by the four scalar masses ($M_{H^\pm},\,M_{h},\,M_{H},\,M_{A}$) and the mixing angle $\tilde{\alpha}$, which are more physical. Clearly $\lambda_2$ must be kept since it does not relate with any of the new parameters. Since $\mu_2$ and $\lambda_3$ always appear in the same combination ($M_{H^\pm}$), we must keep one of them which we choose to be $\lambda_3$. From Eq.~\eqref{eq:mix_ang_lam} we can relate $\lambda_1$ and $\lambda_6$ with $\tilde{\alpha}$, $M_h^2$, $M_H^2$ and $v$. Then we can use Eq.~\eqref{eq:tracematrix} to obtain $\lambda_4$ and $\Sigma$ to obtain $\lambda_5$. With this procedure we
 trade $\mu_2$, $\lambda_1$, $\lambda_4$, $\lambda_5$ and $\lambda_6$ by the four masses and the mixing angle. Our choice for the nine input parameters 
is then: $v,\,M_{H^\pm},\,M_{h},\,M_{H},\,M_{A},\,\tilde{\alpha},\,\lambda_2,\,\lambda_3$ and $\lambda_7$. 
The remaining parameters of the scalar potential can be easily obtained in terms of these inputs:
\begin{align}
\nonumber   & \mu_2=M_{H^\pm}^2-\frac{\lambda_3}{2}v^2\, ,\qquad \lambda_1=\frac{M_h^2+M_H^2\tan^2{\tilde{\alpha}}}{v^2(1+\tan^2{\tilde{\alpha}})}\, ,\qquad \lambda_6=\frac{(M_h^2-M_H^2)\tan{\tilde{\alpha}}}{v^2(1+\tan^2{\tilde{\alpha}})}\, ,\\
    \lambda_4=&\frac{1}{v^2}\left( M_h^2+M_A^2-2M_{H^\pm}^2+\frac{M_H^2-M_h^2}{1+\tan^2{\tilde{\alpha}}}\right) ,\qquad \lambda_5=\frac{1}{v^2}\left( \frac{M_H^2+M_h^2\tan^2{\tilde{\alpha}}}{1+\tan^2{\tilde{\alpha}}}-M_A^2\right) .
\end{align}

While considering interactions of neutral scalars with gauge bosons, $S_1$ plays the role of the SM Higgs boson. This implies that
the couplings of the scalar mass eigenstates with a pair of gauge bosons are given by:
\begin{equation}
g_{hVV}=\cos\tilde{\alpha}\;g_{hVV}^{SM}\, , \qquad\qquad g_{HVV}=-\sin\tilde{\alpha}\;g_{hVV}^{SM}\, , \qquad\qquad g_{AVV}=0\, ,
\label{eq:Higgs_weak_boson_couplings}
\end{equation}
where $VV\equiv W^+W^-, ZZ$.

\subsection{Yukawa Sector}
The interactions of the fermion mass eigenstates with the scalar fields read:
\begin{align}
-\mathcal L_Y\, =\, &\Big(1+\frac{S_1}{v}\Big)\left\{\bar u_L\,M_u\,u_R+\bar d_L\,M_d\,d_R+\bar\nl_L\, M_\nl\,\nl_R\right\}\nonumber\\
&+\frac{1}{v}\, (S_2+i S_3) \left\{\bar u_L\,Y_u\,u_R+\bar d_L\,Y_d\,d_R+\bar\nl_L\, Y_\nl\,\nl_R\right\}\\
&+\frac{\sqrt 2}{v}\, H^+ \left\{\bar u_L\,V\,Y_d\,d_R-\bar u_R \,Y_u^\dagger\,V\,d_L+\bar\nu_L\, Y_\nl\,\nl_R\right\} + \mathrm{h.c.}\, ,\nonumber
\end{align}
where the generation indices are suppressed and the subscripts $L,R$ denote the usual left and right chiral fields. Here, $M_f\;(f\equiv u,d,\nl)$ are diagonal mass matrices for the up-type quark, down-type quarks and charged leptons, respectively, 
originated through the Yukawa interactions with the doublet $\Phi_1$.
$Y_f$ are the Yukawa matrices parametrizing the fermionic couplings with the second doublet $\Phi_2$, and $V$ is the usual CKM matrix required for the quark mixing. In general, 
$Y_f$ could be arbitrary $3\times3$ complex matrices, which leads to unwanted FCNCs at tree level. This can be easily avoided
imposing the alignment condition of $M_f$ and $Y_f$ in flavour space \cite{Pich:2009sp,Pich:2010ic}, i.e.
\begin{equation}\label{eq:Yalignment}
Y_u=\varsigma^*_u\,M_u \qquad\qquad \text{and} \qquad\qquad Y_{d,\nl}=\varsigma_{d,\nl}\,M_{d,\nl}\, ,
\end{equation}
where the Yukawa alignment parameters $\varsigma_f$ could be arbitrary complex numbers. In terms of the scalar and fermion mass eigenstates, the Yukawa Lagrangian takes then the form:
\begin{equation}
-\mathcal L_Y=\sum_{i,f}\Big(\frac{y_f^{\varphi^0_i}}{v}\Big)\,\varphi^0_i\,\Big[\bar f M_f \mathcal{P}_R f\Big]+\Big(\frac{\sqrt 2}{v}\Big) H^+\,\Big[\bar u\,\big\{\varsigma_d V M_d \mathcal{P}_R-\varsigma_u M_u^\dagger V\mathcal{P}_L\big\}\, d+\varsigma_\nl\, \bar \nu M_\nl \mathcal P_R \nl\Big] + \mathrm{h.c.}\, ,
\end{equation}
where $\mathcal P_{L,R}$ are chirality projection operators
and $\varphi_i^0$ 
are the scalar mass eigenstates.

In the following, we will assume a CP-conserving potential and that there are no additional sources of CP violation beyond the CKM phase, i.e. we will only consider real alignment parameters. The Yukawa couplings of the neutral scalars are then given by:
\begin{eqnarray}
&y_{u}^H=-\sin\tilde\alpha+\varsigma_{u}\,\cos \tilde\alpha\, , \qquad\quad y_{u}^h=\cos\tilde\alpha+\varsigma_{u}\,\sin \tilde\alpha\, , \qquad\quad
y_{u}^A= -i\varsigma_{u}\, ,&
\label{eq:Higgs_yuk_up}
\\
&y_{d,\nl}^H=-\sin\tilde\alpha+\varsigma_{d,\nl}\,\cos \tilde\alpha\, , \quad\quad\; \;y_{d,\nl}^h=\cos\tilde\alpha+\varsigma_{d,\nl}\,\sin \tilde\alpha\, , \quad\quad\;  
y_{d,\nl}^A= i\varsigma_{d,\nl}\,.&
\label{eq:Higgs_yuk_down}
\end{eqnarray}


The usual THDM scenarios based on $\mathcal{Z}_2$ symmetries are just particular cases of the more general ATHDM framework; they can be retrieved by imposing $\mu_3=\lambda_6=\lambda_7=0$ along with the following conditions: 
\begin{align}
\label{eq:THDM_types}
&\text{Type I:\;\;} \varsigma_{u}=\varsigma_d=\varsigma_\nl=\cot\beta,\quad \text{Type II:\;\;} \varsigma_{u}=-\frac{1}{\varsigma_d}=-\frac{1}{\varsigma_\nl}=\cot\beta\, ,\quad 
\text{Inert:\;\;}\varsigma_{u}=\varsigma_d=\varsigma_\nl=0\, ,
\nonumber\\
&\text{Type X:\;\;} \varsigma_{u}=\varsigma_d=-\frac{1}{\varsigma_\nl}=\cot\beta 
\qquad \text{and}\qquad
\text{Type Y:\;\;} \varsigma_{u}=-\frac{1}{\varsigma_d}=\varsigma_\nl=\cot\beta \, .
\end{align}
The alignment requirement (\ref{eq:Yalignment}) remains stable under renormalisation when it is protected by $\mathcal{Z}_2$ symmetries \cite{Ferreira:2010xe}. Otherwise, higher-order quantum corrections create a misalignment of $M_f$ and $Y_f$ that generates loop-suppressed FCNC effects. However, the special Yukawa structure of the ATHDM strongly constrain the possible FCNC interactions, making those effects numerically small \cite{Pich:2009sp,Pich:2010ic,Jung:2010ik}. Assuming exact alignment at some high-energy scale (even at the Planck mass), the small misalignment generated by running down to low energies remains well below the current experimental limits \cite{Braeuninger:2010td,Bijnens:2011gd,Penuelas:2017ikk,Gori:2017qwg}.

\section{Fit set up}

\label{sec:set_up}
Our numerical analyses have been performed with the open-source {\tt HEPfit} package \cite{DeBlas:2019ehy}. This code has been widely used,
due to its efficiency and versatility that allows making global fits, both within the SM \cite{deBlas:2021wap} or in general BSM extensions such as the SM effective field theory \cite{Durieux:2019rbz,Miralles:2021dyw} or particular models of new physics  \cite{Eberhardt:2020dat,Eberhardt:2021ebh}, as it is the case of this paper. 
{\tt HEPfit} works within the Bayesian statistics framework and, therefore, we need to choose carefully the priors of our variables. We have a total of ten new degrees of freedom with respect to the SM:
the physical masses of the additional scalars ($M_{H^\pm},\,M_{H}$ and $M_{A}$), the mixing angle of the CP-even neutral scalars ($\tilde{\alpha}$), three quartic couplings of the potential ($\lambda_2$, $\lambda_3$ and $\lambda_7$) and the three Yukawa alignment parameters ($\varsigma_u$, $\varsigma_d$ and $\varsigma_\nl$). Since our fits include all the available information, we have used uniform distributions as priors. In general, the priors are chosen to cover the region of the parameters that is physically relevant. For the mixing angle we have chosen the prior in such a way that at least the 5$\sigma$ region of the posterior probability is contained within the selected range. The quartic couplings are mainly constrained from theory assumptions, which impose a hard cut in the values that these parameters can take. In this case we have chosen a prior wide enough to include all points allowed by the theory constraints. The priors of the Yukawa couplings were also set within the limits allowed by perturbative constraints. 

There is a huge freedom when choosing ranges for the scalar masses. In this analysis we are assuming that the SM Higgs is the lightest scalar, i.e. that 
$M_H$, $M_A$ and $M_{H^\pm}$ are larger than 125 GeV. 
The complementary scenario where the SM Higgs is not the lightest scalar  is phenomenologically very intriguing and capable of displaying very distinct collider signatures at the LHC \cite{Bernon:2014nxa,Bernon:2015qea,Bernon:2015wef}. However, these signatures and the overall phenomenology 
are strongly dependent on the assumed hierarchy of scalar masses
(which BSM particles are considered to be lighter than the SM Higgs). Moreover, a detailed analysis of the different possible scenarios requires the inclusion of additional experimental data from LEP and flavour factories, and a proper theoretical treatment of hadronic resonances in the mass region below 10 GeV.
These intricacies for the light BSM particles are beyond the scope of this paper and therefore, here we focus only on the heavy BSM scenario.\footnote{In a forthcoming publication we will soon address the possibility of lighter scalars.} 

Higher scalar masses are obviously preferred by the data because no clear deviations from the SM have been observed so far.
Taking into account that many direct searches have been studying mass ranges up to 1~TeV, this number seems a reasonable higher cut-off for our global analysis. However, we will also provide some results in which the highest value of the scalar masses have been set to 1.5~TeV, so that we can get a feeling on how much our results depend on the priors. Finally, we would also like to comment that, instead of taking the masses as fundamental parameters, using the masses squared could also be a well-justified choice. However, as can be seen in Ref.~\cite{Eberhardt:2020dat}, when using a uniform distribution for the masses squared the results depend much more on the ranges of masses analysed. Therefore, we have decided to 
impose our priors linearly on the scalar mass parameters.
The chosen priors can be found in Table~\ref{tab:priors}.

\begin{table}[htb]
\begin{center}
\begin{tabular}{|P{1.1cm}|P{1.1cm}|P{1.1cm}|P{1.1cm}|P{1.1cm}|P{1.1cm}|P{1.1cm}|P{1.1cm}|P{1.1cm}|P{1.1cm}|P{1.1cm}|P{1.1cm}|}
\hline
\multicolumn{12}{|c|}{Priors} \\
\hline
\hline
\multicolumn{4}{|c}{$M_{H^\pm} \subset$ [0.125, 1.0\, (1.5)] TeV} & \multicolumn{4}{|c|}{$M_{H} \subset$  [0.125, 1.0\, (1.5)] TeV} & \multicolumn{4}{c|}{$M_{A} \subset$  [0.125, 1.0\, (1.5)] TeV} \\
\hline
\multicolumn{4}{|c}{$\lambda_2 \subset$ [0, 11]} & \multicolumn{4}{|c|}{$\lambda_3 \subset$ [-3, 17]}  & \multicolumn{4}{c|}{$\lambda_7 \subset$ [-5, 5]}  \\
\hline
\multicolumn{3}{|P{3.3 cm}}{$\tilde{\alpha} \subset$ [-0.16, 0.16]} & \multicolumn{3}{|P{3.3 cm}|}{$\varsigma_u \subset$ [-1.5, 1.5]} & \multicolumn{3}{P{3.3 cm}|}{$\varsigma_d \subset$ [-50, 50]} & \multicolumn{3}{P{3.3 cm}|}{$\varsigma_\nl \subset$ [-100, 100]} \\  
\hline
\end{tabular}
\caption{Priors chosen for the new-physics parameters.}
\label{tab:priors}
\end{center}
\end{table}

\section{Fit constraints}
\label{sec:constraints}

We aim to use as much information as we can in order to constrain the parameter space of the ATHDM. We have combined in this work the whole set of theoretical constraints with all relevant experimental limits coming from LHC direct  searches, Higgs data, electroweak precision data, and flavour observables. 

\subsection{Theoretical considerations}

Regarding theoretical constraints, the two conditions that are usually demanded are: a scalar potential bounded from below and perturbative unitarity of the $S$ matrix. The requirement that the scalar potential is bounded from below indicates that it should not go to large negative values, which would make it unstable, for any configuration of the fields. 
For this purpose, it is useful to recast the scalar potential $V$ in the following Minkowskian form \cite{Ivanov:2006yq}:
\begin{equation}
\label{eq:sc_pot_2}
     V=-\,\Mn_\mu\,{\rn}^\mu + \frac{1}{2}\,\Lambda^{\mu}_{\phantom{\mu}\nu}\, \rn_\mu\,\rn^\nu\, ,
\end{equation}
with
\begin{eqnarray}
\quad &\Mn_\mu\, =\, \Big(-\frac{\mu_1+\mu_2}{2},\,-\,\R\,\mu_3, \,\I\,\mu_3,\, -\,\frac{\mu_1-\mu_2}{2}\Big)\, , &
 \nonumber\\  
 &\rn^\mu\, =\, \Big(|\Phi_1|^2+|\Phi_2|^2, \, 2\,\R (\Phi_1^\dagger\Phi_2), \, 2\,\I (\Phi_1^\dagger\Phi_2), \, |\Phi_1|^2-|\Phi_2|^2\Big)\, , 
\end{eqnarray}
and
 \begin{equation}
 \Lambda^{\mu}_{\phantom{\mu}\nu}\, =\,\frac{1}{2}\,\begin{pmatrix}
     \;\frac{1}{2}(\lambda_1+\lambda_2)+\lambda_3 && \R (\lambda_6+\lambda_7) && -\,\I (\lambda_6+\lambda_7) && \frac{1}{2}(\lambda_1-\lambda_2)\\
     - \R (\lambda_6+\lambda_7) && -\lambda_4-\R \lambda_5&& \I \lambda_5 && -\, \R(\lambda_6-\lambda_7)\\
     \I (\lambda_6+\lambda_7) && \I \lambda_5&& -\lambda_4+\R \lambda_5 &&  \I(\lambda_6-\lambda_7)\\
     -\frac{1}{2}(\lambda_1-\lambda_2)&& -\R (\lambda_6-\lambda_7) && \I (\lambda_6-\lambda_7)&&-\frac{1}{2}(\lambda_1+\lambda_2)+\lambda_3\;\;
 \end{pmatrix}\, .
\end{equation}
Diagonalisation of the mixed-symmetric matrix\footnote{Actually $\Lambda_{\mu\nu}$ is a symmetric matrix and can be diagonalized by a $SO(1,3)$ transformation.} $\Lambda_{\phantom{\mu}\nu}^{\mu}$ produces one ``timelike'' ($\Lambda_0$) and three ``spacelike'' ($\Lambda_{1,2,3}$) eigenvalues.\footnote{The attributes ``timelike'' and ``spacelike'' are related to the corresponding eigenvectors.} The scalar potential remains bounded from below 
when the following two conditions are satisfied \cite{Ivanov:2006yq,Ivanov:2015nea}:
\begin{enumerate}
    \item All the eigenvalues are real,
    \item $\Lambda_0>0$ with $\Lambda_0>\Lambda_i$ $\forall\;i\in \{1,2,3\}$.
\end{enumerate}
Several \textit{necessary} conditions for the scalar potential to be bounded from below have been listed in Ref.~\cite{Bahl:2022lio}. However, they are comprehended in the two above-mentioned \textit{necessary and sufficient} conditions.

One additional constraint could be imposed on the quartic couplings by demanding that the vacuum of the scalar potential is a stable neutral minimum.\footnote{This requirement is a bit more restrictive. It can be relaxed by allowing a meta-stable vacuum with a transition time to the true vacuum larger than the age of the universe.} For this purpose one defines the discriminant of the matrix $\xi\, {\mathbb I_4}-\Lambda^{\mu}_{\phantom{\mu}\nu}$ as:
\begin{equation}
    D=-\prod_{k=0}^{3}(\xi-\Lambda_k) \qquad \wi \qquad  \xi=\frac{M_{H^\pm}^2}{v^2}\,,
\end{equation}
 where the Lagrange multiplier $\xi$ is determined by minimizing the scalar potential $V$ with the constraint\footnote{The condition $\rn^\mu\, \rn_\mu=0$ ensures the vacuum to be a charge-neutral minimum. It has been proved that neutral and charge-breaking vacua cannot coexist in any THDM \cite{Ferreira:2004yd,Barroso:2013awa}.}  $\rn^\mu\, \rn_\mu=0$. The condition for a global minimum is found to be \cite{Ivanov:2015nea}: $D>0$, or
$D<0$ with $\xi>\Lambda_0$.
In our analysis we have included this condition, requiring a stable neutral minimum. However, removing this constraint does not change our results significantly.

The unitarity of the $S$ matrix ensures that, in order to conserve the total probability,  scattering amplitudes do not grow monotonically with energy. Since unitarity emerges from the basic formulation of QFT, it is bound to hold in a complete renormalisable theory while dealing with the full $S$ matrix. However, the unitarity bound does not need to be satisfied by the perturbative calculation of the $S$ matrix to any particular order. The stronger requirement of \textit{perturbative unitarity} enforces the unitarity of the $S$ matrix to be obeyed at every order of perturbation theory. It indirectly indicates that the couplings, in terms of which the perturbative expansion is performed, are not very large and one can safely neglect higher-order contributions. Therefore, imposition of perturbative unitarity on all the $2\to2$ scattering amplitudes involving the scalars 
(both the massive ones and the Goldstone bosons, which account for the $W_L^\pm$ and $Z_L$)
will restrict the quartic couplings $\lambda_i$, ensuring that the perturbative expansion of the $S$ matrix does not diverge at high energies. 
\textit{Tree-level unitarity} is enforced on the $2\to2$ scattering matrix of scalars by demanding that:
\begin{equation}
\label{eq:pert}
    (a_j^{0})^2\leq \frac{1}{4} \qquad\text{with}\qquad (\mathbf{a_0})_{i,f}=\frac{1}{16\pi s}\int_{-s}^{0} dt \;\mathcal M_{i\to f}(s,t)\, ,
\end{equation}
where $\mathbf{a_0}$ is the matrix of tree-level partial-wave amplitudes and $a_j^{0}$  the corresponding eigenvalues in the $j^{th}$ partial wave.
Nevertheless, while considering the scattering of scalars at very high energy, only the $S$-wave amplitude with $j=0$ becomes most significant at tree level. 
Constructing two-scalar scattering states with definite hypercharge and weak isospin $(Y,I)$ and grouping the ones with the same set of quantum numbers,
the 25-dimensional\footnote{For any THDM, there are fourteen neutral, eight single-charged and three doubly-charged two-body scalar scattering states possible.} matrix $\mathbf{a_0}$ can be expressed in a block-diagonal form:
\begin{eqnarray}
&\mathbf{a_0^{++}}=\frac{1}{16\pi}\,  X_{(1,1)}\; ,\quad 
\mathbf{a_0^+}=\frac{1}{16\pi}\, \text{diag}\, [X_{(0,1)},\, X_{(1,0)},\, X_{(1,1)}]\; ,&\nonumber\\
&\mathbf{a_0^0}=\frac{1}{16\pi}\, \text{diag}\, [X_{(0,0)},\, X_{(0,1)},\, X_{(1,1)},\, X_{(1,1)}]\; ,&   
\end{eqnarray}
where the superscript indicates the total charge of the initial or final states. Thus the scattering matrix  $\mathbf{a_0}$ is comprised of the following four submatrices \cite{Ginzburg:2005dt,Bahl:2022lio}:
\begin{eqnarray}
\label{eq:pert1}
    &X_{(1,0)}=\lambda_3-\lambda_4\, ,\qquad X_{(1,1)}=\begin{pmatrix}\lambda_1&&\lambda_5&&\sqrt 2 \lambda_6\\
    \lambda_5^*&&\lambda_2&&\sqrt 2 \lambda_7^*\\
    \sqrt 2\lambda_6^*&&\sqrt 2 \lambda_7^*&&\lambda_3+\lambda_4\end{pmatrix} , \qquad X_{(0,1)}=\begin{pmatrix}
        \lambda_1&&\lambda_4&&\lambda_6&&\lambda_6^*\\   \lambda_4&&\lambda_2&&\lambda_7&&\lambda_7^*\\
       \lambda_6^*&&\lambda_7^*&&\lambda_3&&\lambda_5^*\\  
       \lambda_6&&\lambda_7&&\lambda_5&&\lambda_3\\  
    \end{pmatrix},&\nonumber\\
    &X_{(0,0)}=\begin{pmatrix}
        3\lambda_1&&2\lambda_3+\lambda_4&&3\lambda_6&&3\lambda_6^*\\   
        2\lambda_3+\lambda_4&&3\lambda_2&&3\lambda_7&&3\lambda_7^*\\
       3\lambda_6^*&&3\lambda_7^*&&\lambda_3+2\lambda_4&&3\lambda_5^*\\  
       3\lambda_6&&3\lambda_7&&3\lambda_5&&\lambda_3+2\lambda_4\\  
    \end{pmatrix}.&   
\end{eqnarray}
The perturbative-unitarity condition~(\ref{eq:pert}) can be traded in terms of the eigenvalues ($e_i$) of the above four submatrices by demanding:
\begin{equation}
\label{eq:pert2}
    |e_i|< 8\pi.
\end{equation}


On the other hand, considering the coupling of fermions to the charged scalars, we vary the value of $\varsigma_f$ in the perturbative range satisfying $\sqrt{2}\,|\varsigma_f| m_f/v<1$. A more detailed study on the perturbativity of the alignment parameters could be performed based on the procedure mentioned in Ref. \cite{Allwicher:2021rtd}, which would extend the range of $\varsigma_f$ a bit more. However, for our analysis it is sufficient to consider the above inequality.

\subsection{Direct searches}
\label{sec:direct_searches}

Many searches of additional scalars have been developed at the LHC. In order to use those results we have compared the theoretical prediction of the cross-section times branching ratio, $\sigma\cdot \mathcal{B}$, for several processes with the exclusion limits obtained by the CMS and ATLAS experimental collaborations. 

The experimental data are provided in the form of tables, which compile the values of the 95\% upper limits on $\sigma\cdot \mathcal{B}$, as a function of the resonance mass. In these tables, \texttt{HEPfit} performs a linear interpolation if needed.
In order to compare the theoretical results with the experimental data, we define the
ratio of the theoretical prediction over the experimental upper limit. To this ratio \tp{we} 
assign a Gaussian distribution (restricted to positive values) centered at 0 such that the
value 1 is excluded with a 95\% probability.

All the channels included in our fit can be found in Appendix \ref{sec:data_included}. In Tab.~\ref{tab:directsearches4} we show the channels included for the charged scalars, while those of the neutral scalars are shown in Tabs.~\ref{tab:directsearches3}, \ref{tab:directsearches1} and \ref{tab:directsearches2}. Specifically, in Tab.~\ref{tab:directsearches3} are shown the decays into neutral scalars (including the SM Higgs boson); in Tab.~\ref{tab:directsearches1} the decays into fermions, $\gamma\gamma$ and $Z\gamma$; and in Tab.~\ref{tab:directsearches2} the decays into weak gauge bosons. When the limits on $\sigma\cdot \mathcal{B}$ provided in the experimental papers consider also a subsequent decay of the SM particles, we show this final decay with parenthesis. When the limits on $\sigma\cdot \mathcal{B}$ are provided directly on the decay width of the NP particles to SM particles, but using particular decays of the produced SM particles, we show such SM decays with square brackets.

\subsection{Higgs data}

The presence of additional scalars generates relevant effects on the production and decay of the SM Higgs. 
The one-loop $h\to 2\gamma$ amplitude receives additional contributions from the charged scalar.
Furthermore, due to the mixing among the CP-even scalars, the coupling of the SM Higgs with the weak bosons is modified as shown in Eq.~(\ref{eq:Higgs_weak_boson_couplings}), which 
changes the Higgs production through vector boson fusion (VBF) and the associated production with vector bosons (Vh). 
The decays of the Higgs boson to fermions are also sensitive to the scalar mixing, as shown in Eqs.~(\ref{eq:Higgs_yuk_up}) and (\ref{eq:Higgs_yuk_down}), which also modifies the Higgs production through gluon fusion (ggF) and its associated production with $t\bar{t}$ pairs (tth). 

The production and the subsequent decay of the SM Higgs boson have been measured (or bounds have been set) at the LHC for the most relevant production modes (ggF, VBF, Vh and tth) and decay channels ($c\bar{c}$, $b\bar{b}$, $\gamma\gamma$, $\mu^+\mu^-$, $\tau^+\tau^-$, $WW$, $Z\gamma$ and $ZZ$).
These data are parametrised in terms of the Higgs signal strengths, which are defined as the measured cross section times branching ratio for a given production and decay Higgs channel, in units of the SM prediction. Table \ref{Tab:HiggsStrengths}, in Appendix \ref{sec:data_included}, compiles the experimental papers from which we have taken the values of the different Higgs signal strengths relevant to our analysis.

\subsection{Electroweak precision observables}

The presence of additional scalars generates contributions to the oblique parameters (also known as Peskin–Takeuchi parameters \cite{Peskin:1990zt, Peskin:1991sw}) $S$, $T$ and $U$ \cite{Haber:2010bw}. The experimental values of these parameters are obtained from global fits of electroweak precision data, using observables directly measured (mainly) at LEP and SLC. Among those observables, 
we highlight the ratio $R_b\equiv \Gamma(Z\rightarrow b\bar{b})/\Gamma(Z\rightarrow \rm{hadrons})$ \cite{Haber:1999zh,Degrassi:2010ne}, which is also affected by the additional scalars. Using the values of the oblique parameters from the PDG would be inconsistent in our study because those values do not take into account the NP contamination on $R_b$. 
Moreover, since we include $R_b$ directly in our global fit of the ATHDM, the information from this observable cannot be employed to determine the oblique parameters.

In order to obtain non-contaminated values for the oblique parameters, we have repeated the electroweak fit removing $R_b$ from the list of fitted measurements, obtaining the values of $S$, $T$ and $U$ that will be included as inputs to our analysis. Those values are summarised in Tab.~\ref{tab:STU} of Appendix \ref{sec:STU_fit}. 

The values of the oblique parameters are also highly dependent on the value of the $W$-boson mass ($M_W$). For our baseline results on these parameters we have used the value of $M_W$ quoted by the PDG 2022 \cite{ParticleDataGroup:2022pth}. However, in April of 2022 the CDF collaboration
announced a controversial new measurement
of $M_W$ \cite{CDF:2022hxs}, which is incompatible with the SM prediction by 7$\sigma$. Since there is not yet consensus in the community, we have not included this measurement in our main global fit but, instead, we will show how much the results change with the values for the oblique parameters obtained incorporating this new measurement in a global electroweak fit performed with \texttt{HEPfit} \cite{deBlas:2016ojx,deBlas:2022hdk}.\footnote{Note that in these references $R_b$ is also included as an input but we have checked that repeating the fit of Ref.~\cite{deBlas:2022hdk} removing $R_b$ leads to very similar results. Therefore, we decided to use the values quoted in the original paper.}

Finally, we have also noticed that using the three oblique parameters $S$, $T$ and $U$ or fitting only $S$ and $T$ (assuming $U$ to be negligible) gives very similar results. Therefore, we use only $S$ and $T$, since it is well known that the THDM contributions to $U$ are highly suppressed \cite{Haber:2010bw}.

\subsection{Flavour observables}
\label{sec:flavour}
The parameter space of the ATHDM can also be constrained from flavour observables, since the new scalars generate relevant contributions to many of them.
However, in order to be able to determine these contributions, we first need to know the numerical values of the CKM parameters. For this work we have adopted the Wolfenstein parametrisation \cite{Wolfenstein:1983yz}, so we need to provide the corresponding values as inputs in our analysis.
The world averages quoted in the PDG 2022 \cite{ParticleDataGroup:2022pth} originate in SM fits from the CKMfitter \cite{Charles:2004jd} and UTfit \cite{UTfit:2006vpt,UTfit:2022hsi} collaborations, which make use of several flavour transitions that could be affected by the additional scalars.
The UTfit collaboration also provides the values of the Wolfenstein parameters removing the loop processes from the fit \cite{UTfit:2005lis}, which gives the correct values for models in which the tree-level effects of NP are negligible. Nevertheless, for some (small) regions of the parameter space in the ATHDM there could still be some contamination to some of the tree-level processes used to determine the Wolfenstein parameters. For this reason we have decided to repeat this CKM fit, using only processes that are not contaminated by the additional scalars. Details on this fit, as well as the resulting numerical values for the CKM parameters, can be found in Appendix \ref{sec:CKM_fit}. Note also that we use as observables for our ATHDM global fit some processes that are usually taken into account in the CKM fit, like the pseudoscalar-meson leptonic decays, so it is clear that we should remove these processes from our CKM determination.

We will consider all relevant flavour observables that constrain the CP-conserving ATHDM, including the contributions to loop processes 
like the neutral-meson mixing of $B_s$ ($\Delta M_{B_s}$) \cite{Jung:2010ik,Chang:2015rva}, the weak radiative decay $B\rightarrow X_s \gamma$ \cite{Bobeth:1999ww,Misiak:2006ab,Misiak:2006zs,Jung:2010ik,Jung:2010ab,Jung:2012vu,Hermann:2012fc, Misiak:2015xwa, Misiak:2017woa, Misiak:2020vlo} and the rare weak leptonic decay $B_s\rightarrow \mu^+\mu^-$ \cite{Li:2014fea,Arnan:2017lxi}. Besides these loop processes, we have also included some relevant tree-level transitions like the leptonic decays of heavy pseudoscalar mesons ($B\rightarrow \tau \nu$, $D_{(s)}\rightarrow \mu \nu$ and $D_{(s)}\rightarrow \tau \nu$), as well as the ratios of the leptonic  decays of light pseudoscalar mesons ($\Gamma(K\rightarrow \mu\nu)/\Gamma(\pi\rightarrow \mu\nu)$) and the similar tau decays ($\Gamma(\tau\rightarrow K\nu)/\Gamma(\tau\rightarrow \pi\nu)$) \cite{Jung:2010ik}. 

We have not included in our main fit the anomalous magnetic moment of the muon, $(g-2)_\mu$ \cite{Ilisie:2015tra},
 because there is no full consensus in the community on the SM prediction of this observable. The most recent lattice computations of the hadronic vacuum polarisation \cite{Borsanyi:2020mff,Ce:2022kxy,ExtendedTwistedMass:2022jpw} and the estimates from $\tau$ decay
\cite{Davier:2010fmf,Pich:2013lsa} seem to find a SM result much closer to the experimental value \cite{Muong-2:2021ojo,Muong-2:2023cdq} than the dispersive $e^+e^-$ prediction \cite{Aoyama:2020ynm}, and additional hints of a possible dispersive underestimate are suggested by a recent QCD analysis of the Adler function \cite{Davier:2023hhn} and the CMD3 data \cite{CMD-3:2023alj}.
Nevertheless, this observable is included in our code and we will show the individual constraints obtained with it, comparing those results with the ones obtained using all the other flavour observables that we have mentioned here.

\section{Fit results}
\label{sec:results}

Here we provide the results for all the fits we have performed in this analysis. We discuss first in several subsections the limits obtained using only some subsets of observables, in order to show the relevance that each observable has in the global fit. 
Afterwards, we present
the global fit, which provides the main results of this 
work. We
show also how the results of our global fit would change if
the new CDF measurement of $M_W$ \cite{CDF:2022hxs} is included.

\subsection{Theoretical bounds}

The theoretical considerations generate constraints on the parameters of the scalar potential.
As shown in Section \ref{sec:Model}, we have a total of nine degrees of freedom in the CP-conserving potential.
The scalar vacuum expectation value and $M_h$
are fixed by the measurement of the Fermi constant in muon decay \cite{ParticleDataGroup:2022pth} and the Higgs mass measurement at the LHC \cite{ParticleDataGroup:2022pth}. We then have a total of seven degrees of freedom which could be possibly constrained with the theory assumptions. As mentioned before, we 
trade some of these scalar potential parameters 
with more physical inputs. Our chosen parameters are the masses of the new scalars, the mixing angle among the CP-even scalars, and the three quartic couplings $\lambda_2$, $\lambda_3$ and $\lambda_7$. 

\begin{figure}[h!]
    \centering
    \hspace*{-0.8 cm}
    \includegraphics[scale=0.42]{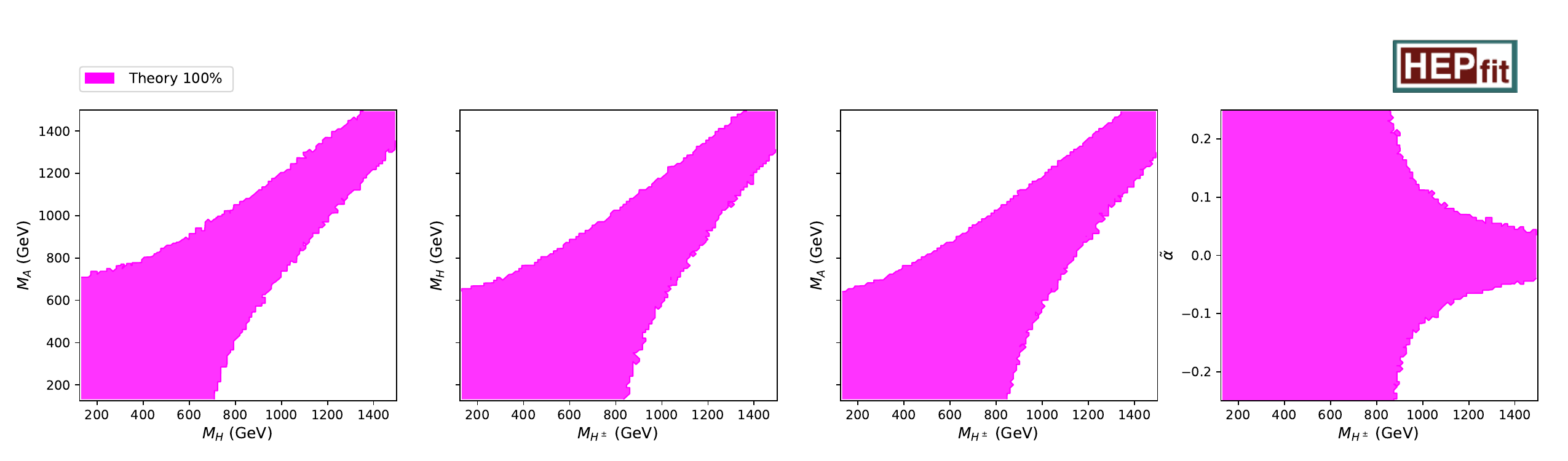}
    \caption{Constraints from theory assumptions.}
    \label{fig:theo_masses}
\end{figure}

The theoretical constraints on the parameters of the potential can be translated into limits on the scalar mass splittings.
Fig.~\ref{fig:theo_masses} displays the resulting  constraints on the correlation among the different masses and the correlation of the charged scalar mass with the mixing angle. Note that in this case we are showing the 100\% probability region, since all the points that satisfy these constraints are equally valid. In general, we observe that for scalar mass values below 700 GeV, the constraints on the mass splittings and
the mixing angle are rather weak. This can be easily understood, since for low masses the mass differences cannot be large enough to reach the theoretical bounds.
However, above this threshold, especially beyond 1 TeV, the
constraints become considerably more stringent. 
Since differences of masses squared are proportional to combinations of quartic couplings  times the vev squared, the naive estimate $\sqrt{4\pi} v\approx 870$ GeV gives indeed a good idea of the scale where these limits become relevant.

\begin{figure}[h!]
    \centering
    \hspace*{-0.5 cm}
    \includegraphics[scale=0.49]{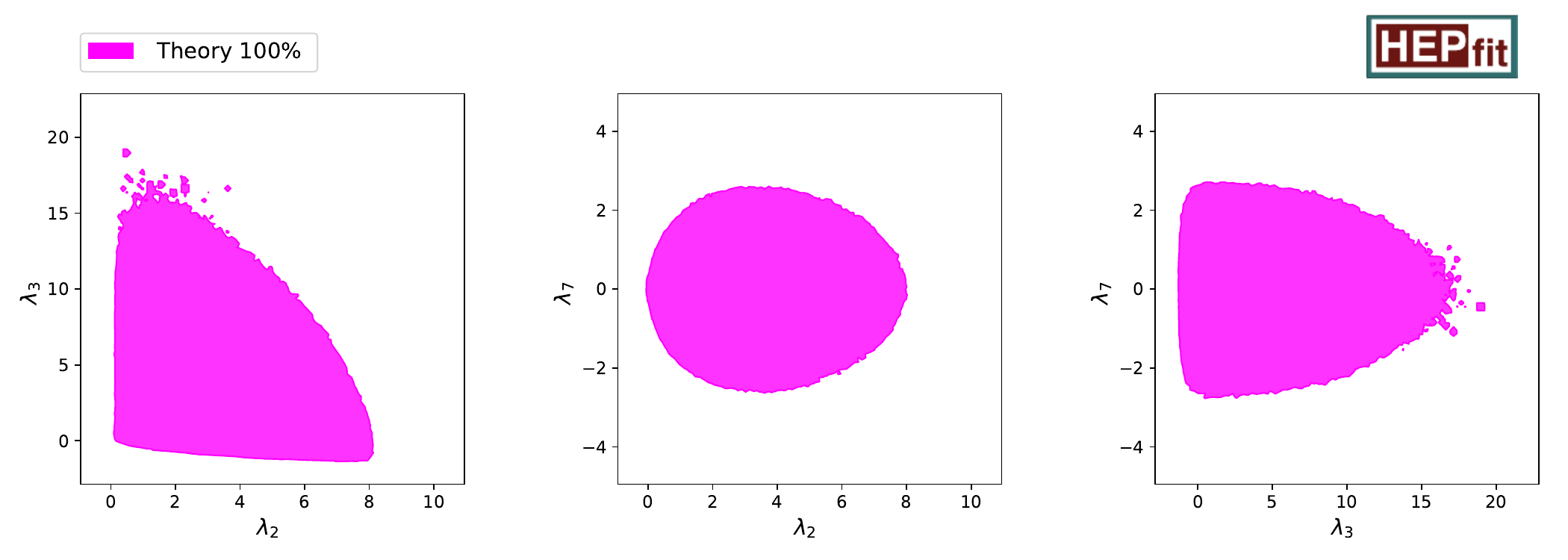}
    \caption{Constraints from theory assumptions.}
    \label{fig:theo_lambdas}
\end{figure}

\FloatBarrier

The constraints on the quartic couplings are shown in Fig.~\ref{fig:theo_lambdas}, where we can see that negative values of $\lambda_2$ are forbidden while slightly negative values of $\lambda_3$ could be allowed, provided they are
very close to zero. Finally, $\lambda_7$ is clearly constrained to be smaller than 3 in modulus.


\subsection{Experimental bounds}

\begin{figure}[h!]
    \centering
    \hspace*{-1.3 cm}
    \includegraphics[scale=0.49]{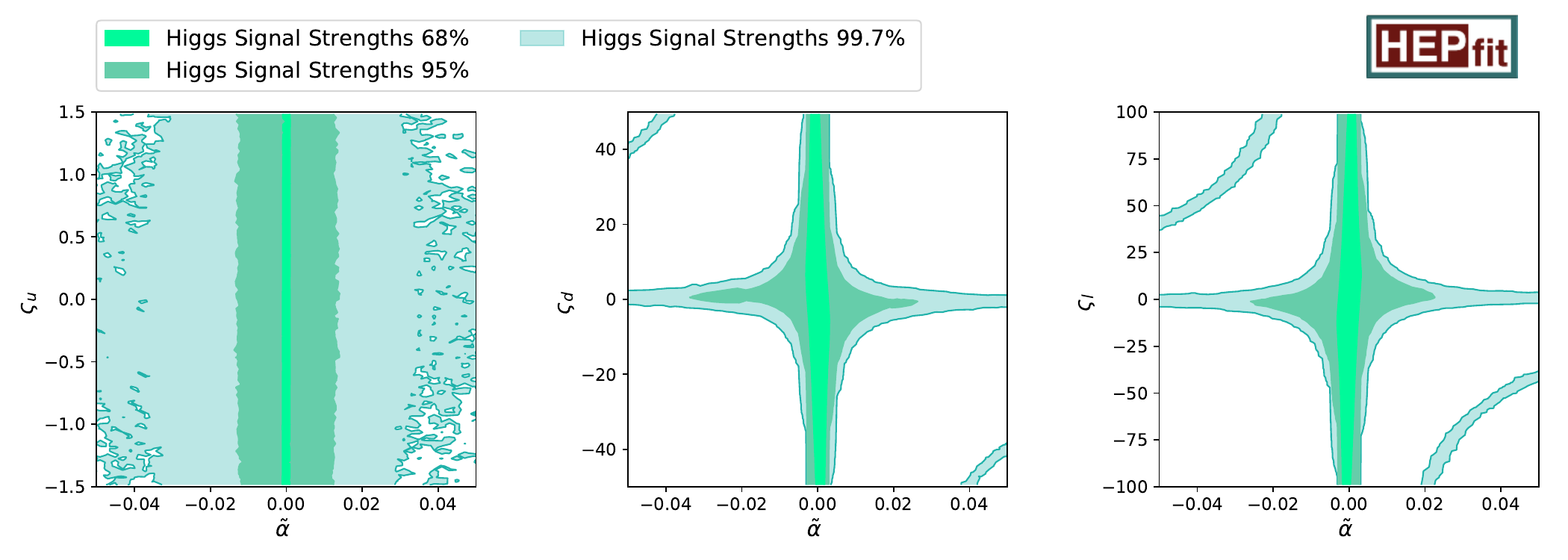}
    \caption{Constraints from the measured Higgs signal strengths.}
    \label{fig:sig_str}
\end{figure}

Fig.~\ref{fig:sig_str} shows that the measured Higgs signal strengths generate tight constraints in the 
$\tilde\alpha - \varsigma_f$ planes.
At 99.7\% probability some allowed regions, at the corners of the central and right plots of Fig.~\ref{fig:sig_str}, appear really far away from the SM solution. 
These regions correspond to down-type and/or lepton Yukawas of opposite sign to the SM Higgs couplings.
Since these observables are sensitive only to the modulus of the Yukawa couplings, this could be a possible solution, although following a bayesian approach the prior probability of this kind of solutions would be quite small. 
No such regions appear for up-type quarks because the relative sign between the top Yukawa and the Higgs coupling to the gauge bosons gets constrained by the $h\to 2\gamma$ decay width and the range of allowed $|\varsigma_u|$ values is much smaller.

\begin{figure}[h!]
    \centering
    \hspace*{-1.5 cm}
    \includegraphics[scale=0.5]{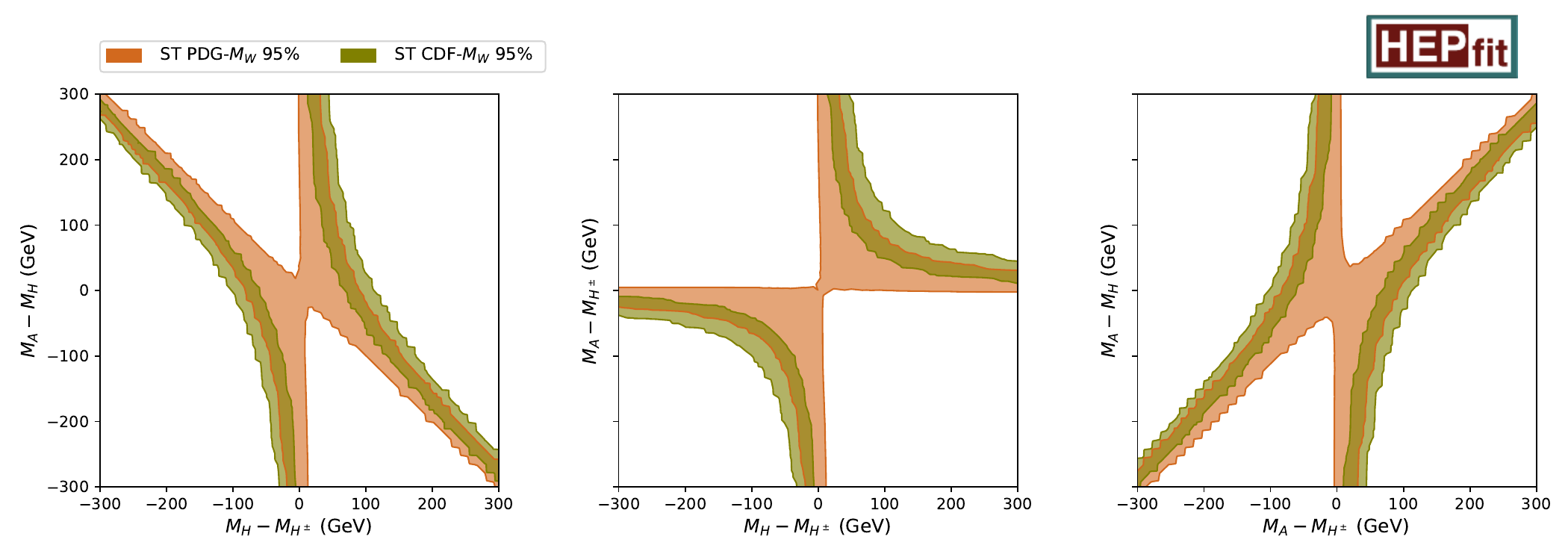}
    \caption{Constraints on the mass splittings from the oblique parameters $S$ and $T$.}
    \label{fig:ST}
\end{figure}

The oblique parameters ($S$ and $T$) generate significant constraints on the mass splitting of the additional scalars. These constraints can be found in Fig.~\ref{fig:ST}, where we
compare the results obtained when $S$ and $T$ are fitted with the PDG 2022 value of $M_W$ (PDG-$M_W$) with those
using the modified $S$ and $T$ values emerging from the world average~\cite{deBlas:2022hdk} after including the CDF $M_W$ (CDF-$M_W$) measurement \cite{CDF:2022hxs}. 
The allowed regions using the PDG value of $M_W$ are compatible with small or even zero mass splittings. However, once the CDF value is included in the fit, a non-zero mass difference between the new scalars is definitely required to explain the deviation from the SM prediction of $M_W$. This scalar mass difference could be accommodated within this model, since we are able to generate the needed contribution to $M_W$, even when we impose agreement with all other observables in combination, as we will discuss in Section~\ref{sec:CDF-MW}.

\begin{figure}[h!]
    \centering
    \hspace*{-1.3 cm}
    \includegraphics[scale=0.49]{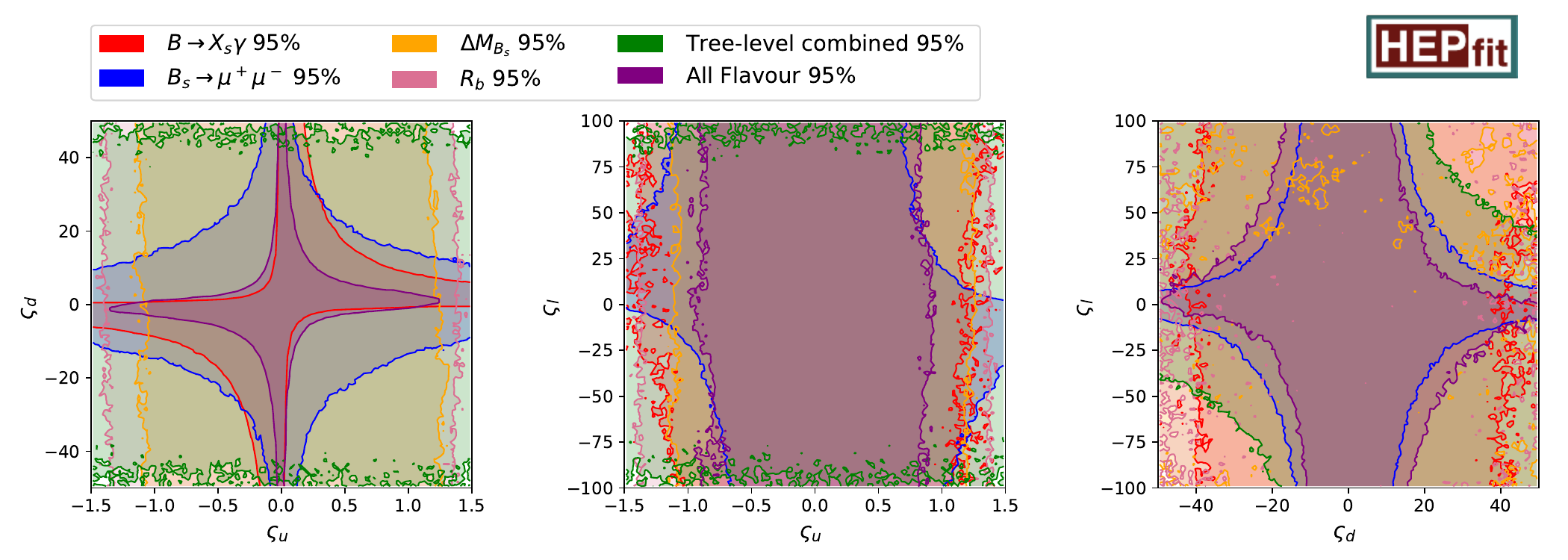}
    \caption{Constraints from flavour observables on the alignment parameters. }
    \label{fig:flav_plot_couplings}
\end{figure}

\begin{figure}[h!]
    \centering
    \hspace*{-1.3 cm}
    \includegraphics[scale=0.49]{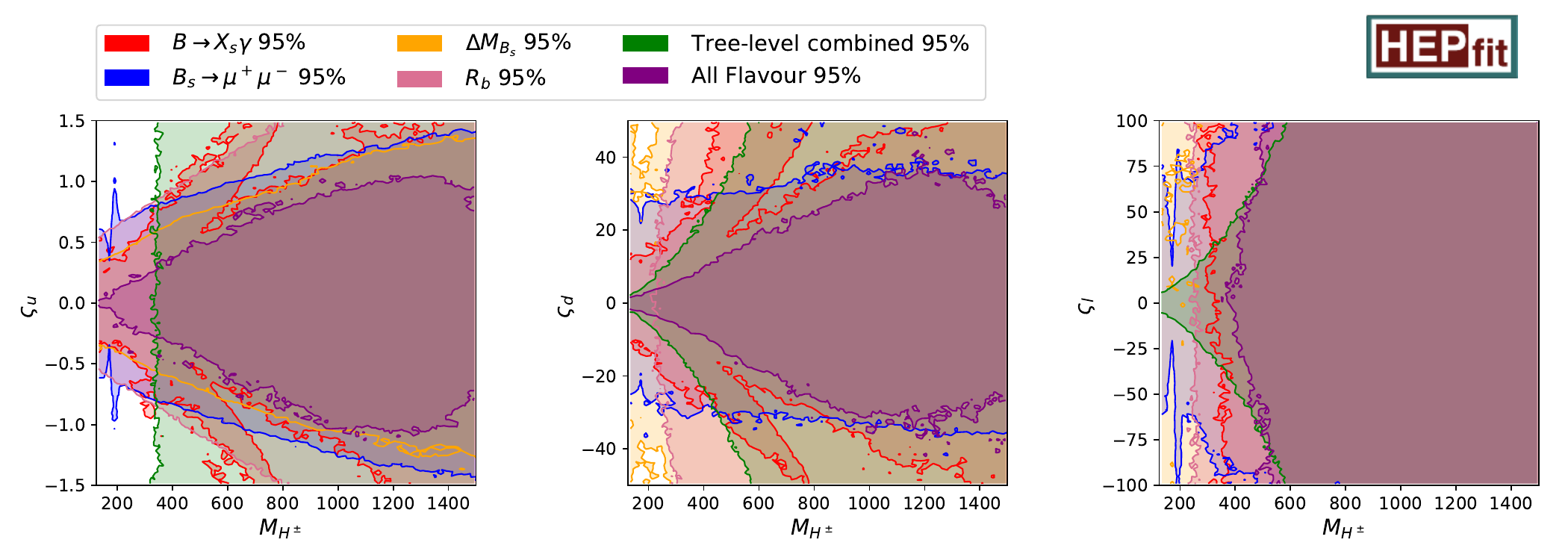}
    \caption{Constraints from flavour observables on the Yukawa couplings and mass of the charged scalar. }
    \label{fig:flav_plot_couplings_vs_mass}
\end{figure}

\begin{figure}[h!]
    \centering
    \hspace*{-1.3 cm}
    \includegraphics[scale=0.49]{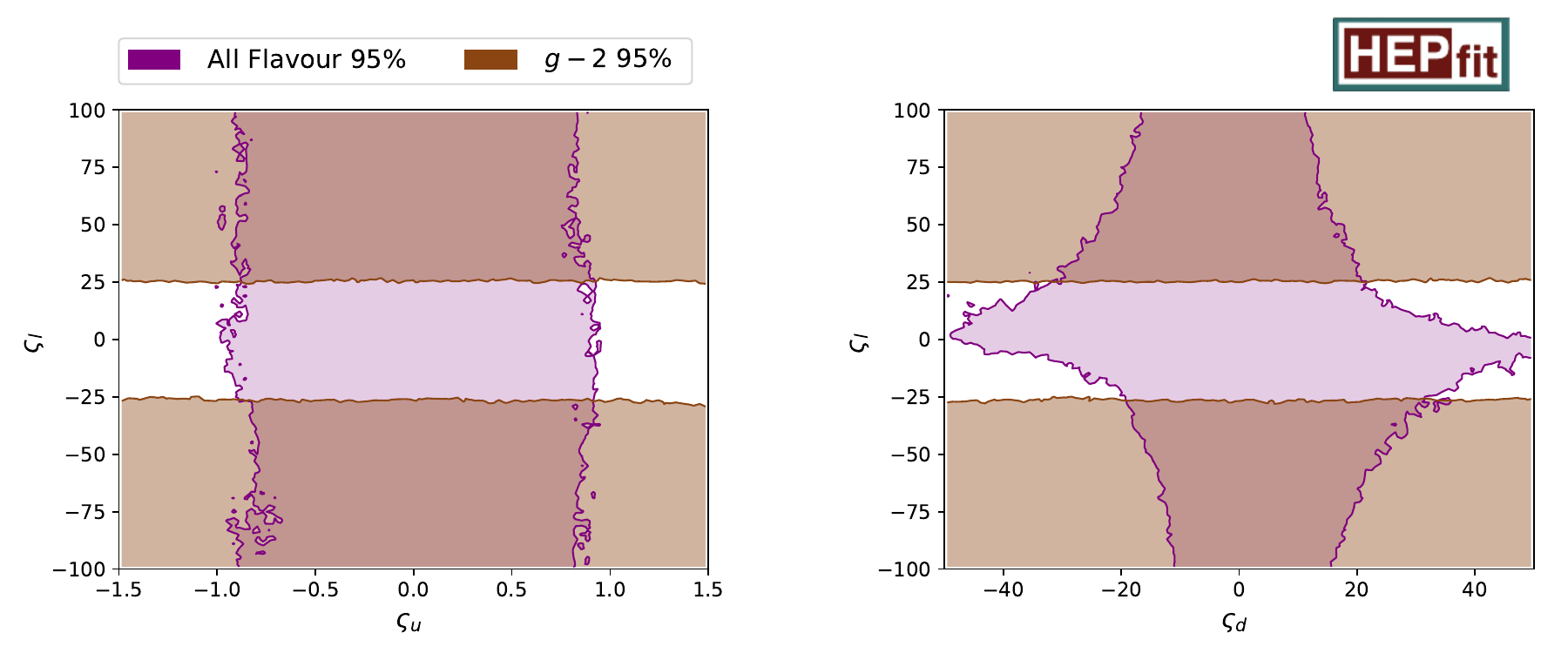}
    \caption{Comparison of the $\varsigma_\nl$ region required to accommodate $(g-2)_\mu$ with the flavour constraints.}
    \label{fig:flav_gm2}
\end{figure}

The flavour observables and $R_b$ generate
constraints on the Yukawa alignment parameters ($\varsigma_u$, $\varsigma_d$, $\varsigma_\nl$) and the mass of the charged Higgs $M_{H^\pm}$.
Fig.~\ref{fig:flav_plot_couplings} displays the correlations among $\varsigma_u$, $\varsigma_d$ and $\varsigma_\nl$, while in Fig.~\ref{fig:flav_plot_couplings_vs_mass} we show the correlation of those parameters with $M_{H^\pm}$. 
The dominant NP contribution to the $B\to X_s\gamma$ amplitude is proportional to the product $\varsigma_u\varsigma_d$, which explains the strong correlation between these two alignment parameters observed in the left panel of Fig.~\ref{fig:flav_plot_couplings}.
The first two panels in Fig.~\ref{fig:flav_plot_couplings_vs_mass} exhibit also the presence of two additional bands emerging from the central 
$B\to X_s\gamma$ allowed region; they correspond to solutions with a NP contribution equal to minus two times the SM amplitude.
The strongest constraint on the $\varsigma_u-M_{H^\pm}$ plane comes from meson mixing, although $R_b$, $B\rightarrow X_s\gamma$ and $B_s\rightarrow \mu^+\mu^-$ also help in constraining this plane.
In the $\varsigma_d-M_{H^\pm}$ plane the tree-level 
leptonic and semileptonic decays of pseudoscalar mesons (and tau decays into pseudoscalars) generate the strongest limits
for low values of the scalar masses. However, the tree-level NP contributions rapidly drop when increasing the mass of the charged scalar, so for larger mass values the loop processes dominate. $B\rightarrow X_s\gamma$ provides the strongest limits for intermediate values of $M_{H^\pm}$ around 500 GeV,
while for heavier masses $B_s\rightarrow \mu^+\mu^-$ produces the most relevant constraints because
it also gets contributions from neutral scalar exchanges that are sizeable at large $|\varsigma_d|$.
In the $\varsigma_\nl-M_{H^\pm}$ plane the constraints are very poor, except at very low values of $M_{H^\pm}$ where useful limits are obtained from the tree-level processes.
As shown in the first two panels of Fig.~\ref{fig:flav_plot_couplings_vs_mass}, small values of $M_{H^\pm}$ are allowed when $\varsigma_{u,d}\to 0$, but when marginalising over these two alignment parameters their probability becomes very small, which explains the lower bound on the charged scalar mass observed in the third panel.
Notice that the maximum and minimum values allowed for the different parameters are slightly different in Figs.~\ref{fig:flav_plot_couplings} and \ref{fig:flav_plot_couplings_vs_mass}; this is just a consequence of having non-Gaussian distributions.

Finally, in Fig.~\ref{fig:flav_gm2} we compare the constraints on $\varsigma_\nl$ obtained from the combination of all flavour observables with the results needed to explain the anomalous magnetic moment of the muon. Since we are only considering scalar masses above 125~GeV, quite large values of the leptonic alignment parameter are required, but
there is, indeed, room to satisfy all the flavour constraints in combination \tp{with} $(g-2)_\mu$.
Nevertheless, we decided to exclude this observable from our global fit 
because its SM prediction is currently under debate, as discussed in Section \ref{sec:flavour}.

\FloatBarrier

\subsection{Global Fits}

\label{sec:globa_fits_results}

\begin{table}[h!]
\begin{center}
\begin{tabular}{|P{1.1cm}|P{1.1cm}|P{1.1cm}|P{1.1cm}|P{1.1cm}|P{1.1cm}|P{1.1cm}|P{1.1cm}|P{1.1cm}|P{1.1cm}|P{1.1cm}|P{1.1cm}|}
\hline
\multicolumn{12}{|c|}{\bf Marginalised Individual results} \\
\hline
\hline
\multicolumn{12}{|c|}{ \it Masses up to 1 TeV} \\
\hline
\multicolumn{4}{|P{4.4 cm}}{$M_{H^\pm} \ge$ 390 GeV} & \multicolumn{4}{|P{4.4 cm}|}{$M_{H} \ge$ 410 GeV} & \multicolumn{4}{P{4.4 cm}|}{$M_{A} \ge$ 370 GeV} \\
\hline
\multicolumn{4}{|P{4.4 cm}}{$\lambda_2$: $3.2 \pm 1.9$} & \multicolumn{4}{|P{4.4 cm}|}{$\lambda_3$: $5.9 \pm 3.5$}  & \multicolumn{4}{P{4.4 cm}|}{$\lambda_7:$ $0.0 \pm 1.1$}  \\
\hline
\multicolumn{3}{|P{3.8 cm}}{$\tilde{\alpha}:$ $(0.05 \pm 21.0)\cdot10^{-3}$} & \multicolumn{3}{|P{3.3 cm}|}{$\varsigma_u:$ $0.006 \pm 0.257$} & \multicolumn{3}{P{3.3 cm}|}{$\varsigma_d:$ $0.12 \pm 4.12$} & \multicolumn{3}{P{3.3 cm}|}{$\varsigma_\nl:$ $-0.39 \pm 11.69$} \\  
\hline
\hline
\multicolumn{12}{|c|}{ \it Masses up to 1.5 TeV} \\
\hline
\multicolumn{4}{|P{4.4 cm}}{$M_{H^\pm} \ge$ 480 GeV} & \multicolumn{4}{|P{4.4 cm}|}{$M_{H} \ge$ 490 GeV} & \multicolumn{4}{P{4.4 cm}|}{$M_{A} \ge$ 480 GeV} \\
\hline
\multicolumn{4}{|P{4.4 cm}}{$\lambda_2$: $3.2 \pm 1.9$} & \multicolumn{4}{|P{4.4 cm}|}{$\lambda_3$: $5.9 \pm 3.8$}  & \multicolumn{4}{P{4.4 cm}|}{$\lambda_7:$ $0.0 \pm 1.2$}  \\
\hline
\multicolumn{3}{|P{3.8 cm}}{$\tilde{\alpha}:$ $(0.8 \pm 16.8)\cdot10^{-3}$} & \multicolumn{3}{|P{3.3 cm}|}{$\varsigma_u:$ $-0.011 \pm 0.407$} & \multicolumn{3}{P{3.3 cm}|}{$\varsigma_d:$ $-0.096 \pm 6.22$} & \multicolumn{3}{P{3.3 cm}|}{$\varsigma_\nl:$ $-1.18 \pm 17.54$} \\  
\hline
\end{tabular}
\caption{Marginalised individual results. The mass limits are at 95\% probability while for the others we show the mean value and the square root of the variance. }
\label{tab:marginalised_results}
\end{center}
\end{table}

In this section we provide the results emerging from the global fit, using all the experimental observables and theoretical considerations at the same time. As mentioned in Section~\ref{sec:set_up}, we have considered masses up to 1 and 1.5 TeV, in such a way that we can also explicitly see the dependence on the adopted priors. The marginalised probabilities, given
in Tab.~\ref{tab:marginalised_results},
exhibit indeed
some dependence on the priors, specially regarding the mass limits. This behaviour is reasonable since the likelihood is maximised for higher values of the masses; 
smaller masses would always be disfavoured, moving the 95\% region further away when we allow for higher mass values. The allowed ranges of the Yukawa alignment parameters are also wider when heavier masses are allowed because the scalar contributions to the flavour observables decrease with increasing mass values and, therefore, higher values of 
$\varsigma_{u,d,\nl}$ become possible,
as can be observed in Fig.~\ref{fig:flav_plot_couplings_vs_mass}. 
In contrast, the preferred region of the mixing angle shrinks when we scan over heavier masses because the theoretical constraints, shown in Fig.~\ref{fig:theo_masses}, become more severe for higher values of the scalar masses.


\begin{figure}[h!]
    \centering
    \hspace*{-1.35 cm}
    \includegraphics[scale=0.485]{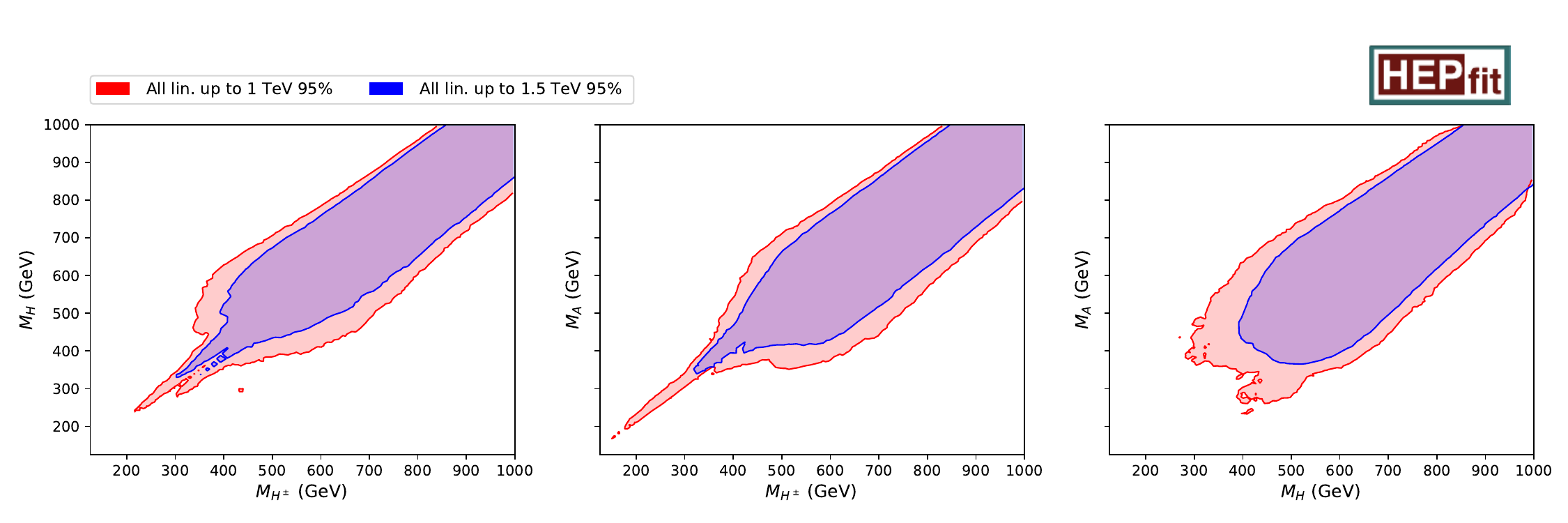}
    \caption{Correlations among the masses of the new scalars.}
    \label{fig:All_Masses}
\end{figure}

\begin{figure}[h!]
    \centering
    \hspace*{-1.3 cm}
    \includegraphics[scale=0.51]{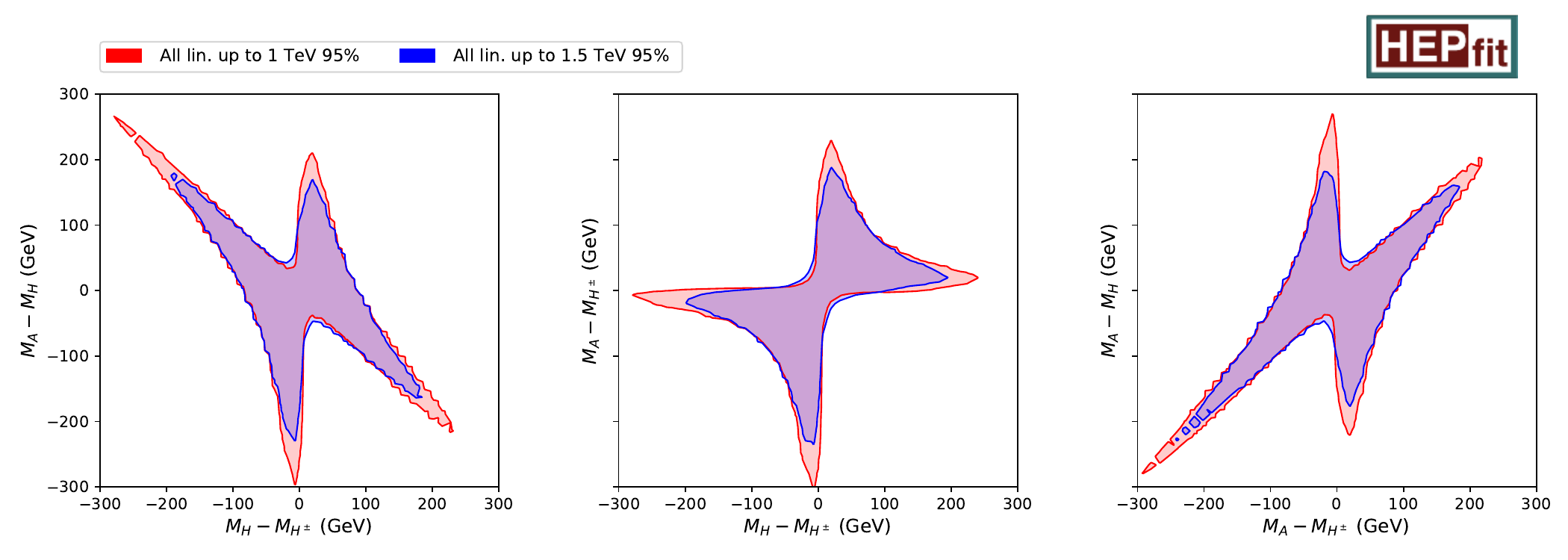}
    \caption{Correlations among the mass splittings.}
    \label{fig:Delta_mass_vs_delta_mass}
\end{figure}

\begin{figure}[h!]
    \centering
    \hspace*{-1.0 cm}
    \includegraphics[scale=0.46]{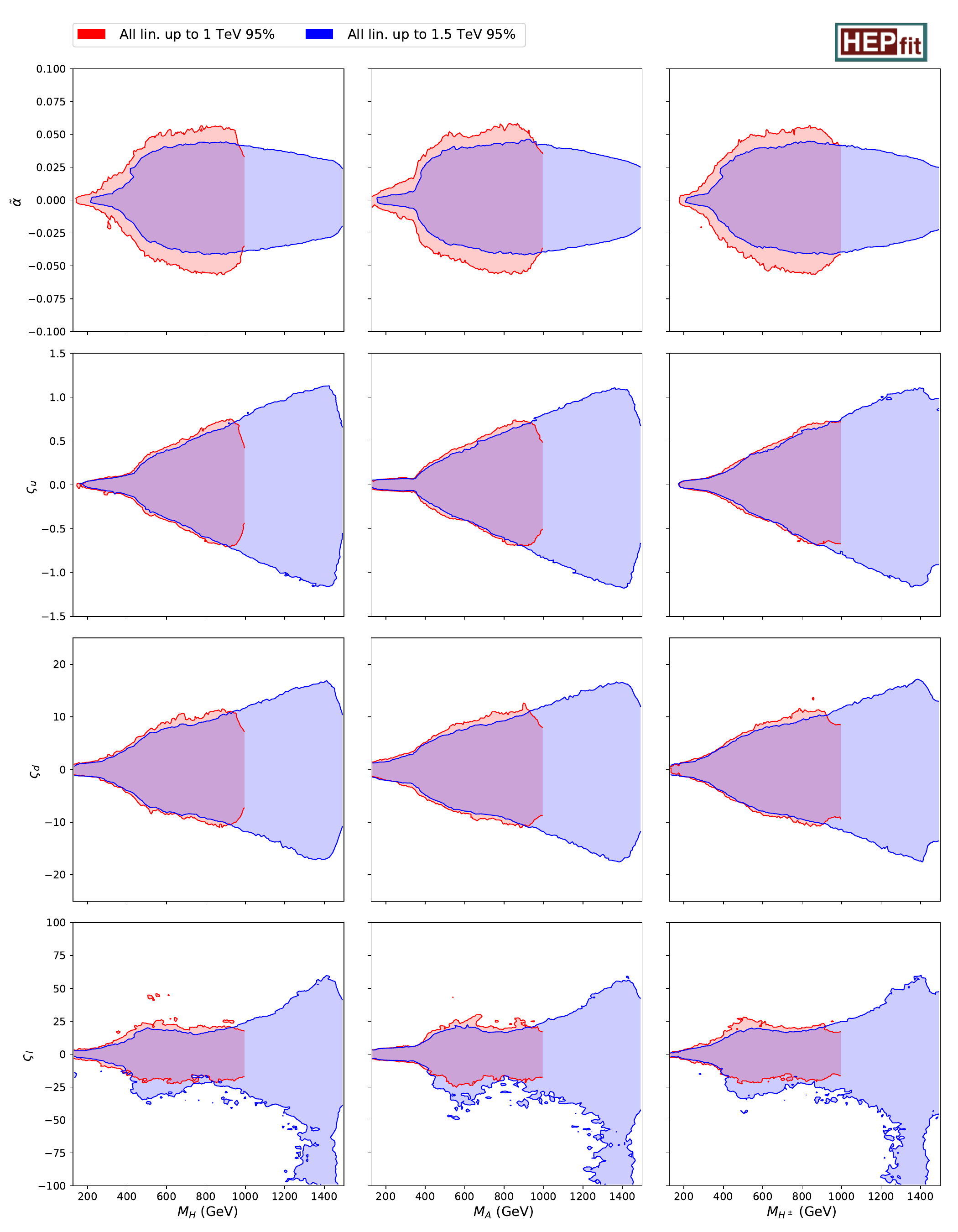}
    \caption{Correlations among the mixing angle and the Yukawa alignment parameters with the masses.}
    \label{fig:Masses_vs_alpha_and_couplings}
\end{figure}

The correlations among the scalar masses are
shown in Fig.~\ref{fig:All_Masses}, 
which exhibits diagonal bands enforced by the theoretical constraints and the oblique parameters. The bounds on the masses get significantly relaxed, getting close to disappear, 
in the limit of zero mass splittings between the charged and the neutral scalars.
This is due to the reduction on the constraints from the oblique parameters, which makes these points much more favoured. 
The correlations among the mass splittings are shown in Fig.~\ref{fig:Delta_mass_vs_delta_mass}. 
In the middle plot one observes that the two mass differences between the charged scalar and the neutral CP-even and CP-odd bosons cannot be simultaneously large; when one increases the other must decrease.
From the left and right plots we can also see that when the mass splitting among the neutral scalars gets large the mass splitting among the charged scalar and one of the neutral scalars must go to zero.  Here again, 
this behaviour is generated by
the oblique parameters, whose NP contribution tends to zero when the mass splitting between any neutral scalar and the charged scalar vanishes. The last interesting plots regarding the masses are their correlations with
the Yukawa alignment parameters and the mixing angle, shown in Fig.~\ref{fig:Masses_vs_alpha_and_couplings}. Here we can see how the higher the masses get, the higher the Yukawa alignment parameters can be. Nevertheless, for masses above 700 GeV, smaller values of the mixing angle are preferred due to the theoretical constraints, as explained before. Finally, for very small values of the Yukawa alignment parameters or the mixing angle the limits on the masses also start to disappear, as expected.

\begin{figure}[h!]
    \centering
    \hspace*{-1.0 cm}
    \includegraphics[scale=0.45]{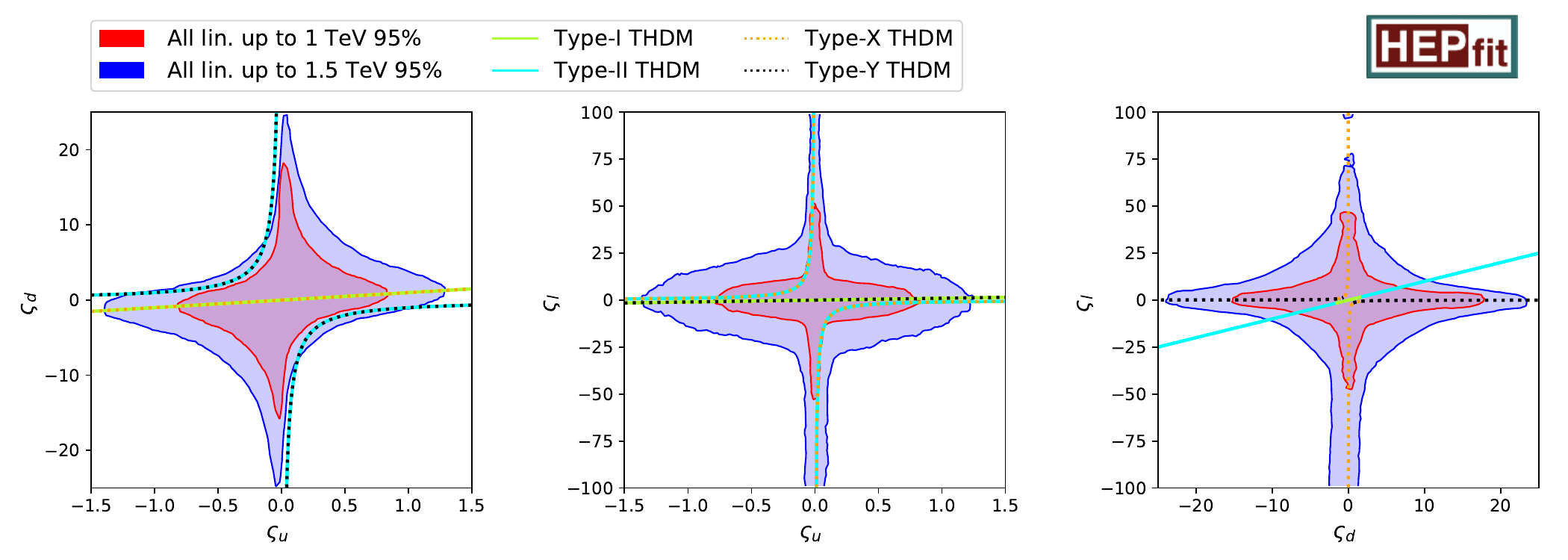}
    \vspace*{-0.3cm}
    \caption{Correlation among the Yukawa alignment parameters.
    Also shown are the restrictions (\ref{eq:THDM_types}) imposed in the different $\mathcal{Z}_2$ THDMs (the inert is omitted since in that case these couplings are simply set to zero).}
    \label{fig:Couplings}
\end{figure}

\begin{figure}[h!]
    \centering
    \hspace*{-1.0 cm}
    \includegraphics[scale=0.45]{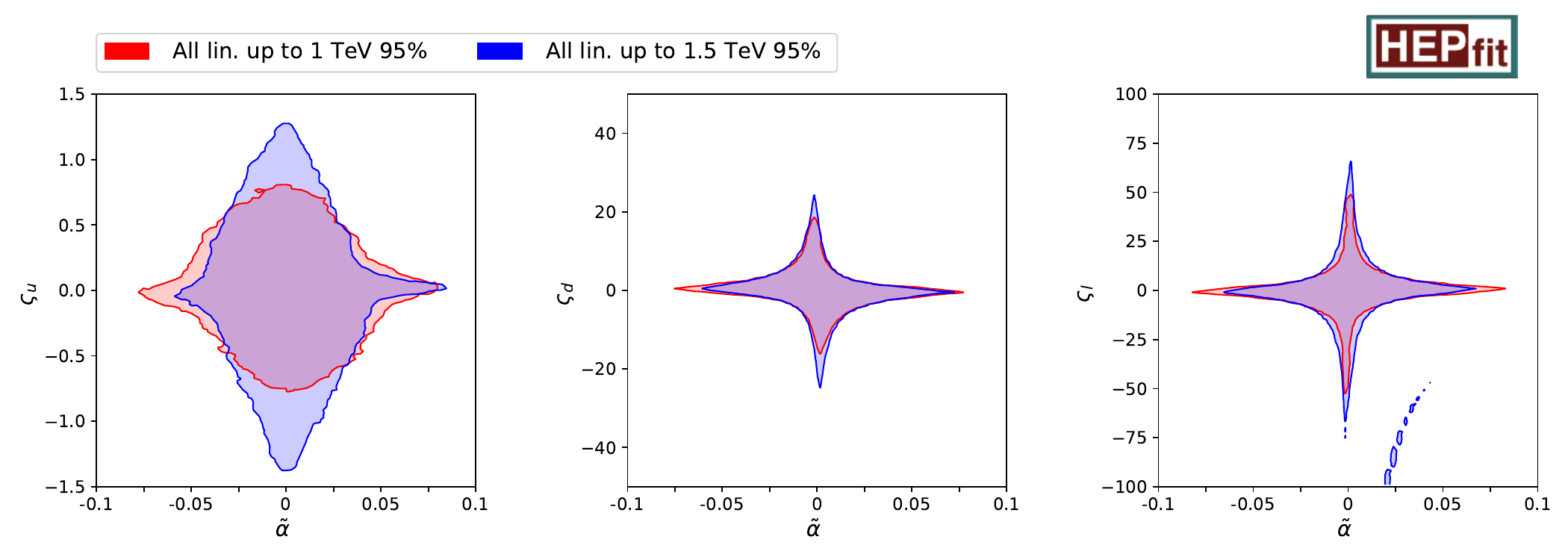} 
    \vspace*{-0.3cm}
    \caption{Correlation among the mixing angle and the Yukawa alignment parameters}
    \label{fig:Couplings_vs_alpha}
\end{figure}

The correlations of the Yukawa alignment parameters 
among themselves and with the mixing angle are shown in Figs.~\ref{fig:Couplings} and \ref{fig:Couplings_vs_alpha}, respectively. In the first one we observe that the regions get wider when higher masses are allowed, since all experimental constraints become softer,
as also shown in Fig.~\ref{fig:Masses_vs_alpha_and_couplings}. 
Moreover, once one of the Yukawa alignment parameters tends to high values, the others get highly suppressed. Indeed, for masses up to 1 TeV, in order to obtain values of $\varsigma_d$ or $\varsigma_\nl$ of order 10, $\varsigma_u$ must be of order 0.1, or smaller.  We observe a similar behaviour looking at the correlations of the Yukawa alignment parameters with the mixing angle. The higher the mixing angle, the smaller the Yukawa alignment parameters must be; and the higher the allowed range of masses, the higher the Yukawa alignment parameters can be. However, as can also be seen in Fig.~\ref{fig:Masses_vs_alpha_and_couplings}, the mixing angle does not grow with the masses, 
and slightly smaller values of $\tilde\alpha$ are preferred when heavier masses are allowed in the fit.
Nevertheless, 
performing a fit with a wider mass region does not significantly change the shape of the correlations between the Yukawa alignment parameters and the mixing angle with respect to the masses of the scalars. 
Integrating
over the masses, like in Figs.~\ref{fig:Couplings} and \ref{fig:Couplings_vs_alpha}, 
the shape stays basically unchanged and just rescales when a different mass range is covered. Indeed, 
performing the fits for these two different mass ranges allows us to get an idea of the shape of the multidimensional correlation of the masses against the other parameters which, unfortunately, cannot be printed in a two dimensional plot. 

In Fig.~\ref{fig:Couplings} we also show the restrictions (Eq. \eqref{eq:THDM_types}) that each $\mathcal{Z}_2$ THDM would impose on the Yukawa alignment couplings. 
In these models all alignment parameters are related
through an angle $\beta$,  whose tangent measures the ratio of the vevs  of the  two scalar fields in the basis where the discrete $\mathcal{Z}_2$ symmetry is imposed. 
Therefore, each $\mathcal{Z}_2$-symmetric model corresponds to
specific curves (either straight lines or hyperbolas)
in the $\varsigma_{f_1}^{}-\varsigma_{f_2}^{}$ planes, 
each point on which represents different values of $\tan \beta$. 
The curves in green, cyan, orange and black depict the Type-I, Type-II, Type-X and Type-Y THDMs, respectively,
whereas the inert THDM corresponds to an isolated point at the origin.
The first panel in Fig.~\ref{fig:Couplings} indicates that the type-II and type-Y models have some tension with the (mainly flavour) data. 
Note, however, that this is just a qualitative comparison. In order to analyse the parameter space of any of these particular models, we would need to change the priors of our analysis by considering $\tan \beta$ as the free parameter instead of the alignment parameters along with the constraint $\mu_3=\lambda_6=\lambda_7=0$.
A comprehensive investigation of the parameter space for each of these models, though valuable, falls outside the scope of this study. 
We refer the interested reader to the more recent specific analyses in Refs. \cite{Chowdhury:2017aav, Haller:2018nnx}.

\begin{figure}[h!]
	\centering
	\hspace*{-1.0 cm}
	\includegraphics[scale=0.45]{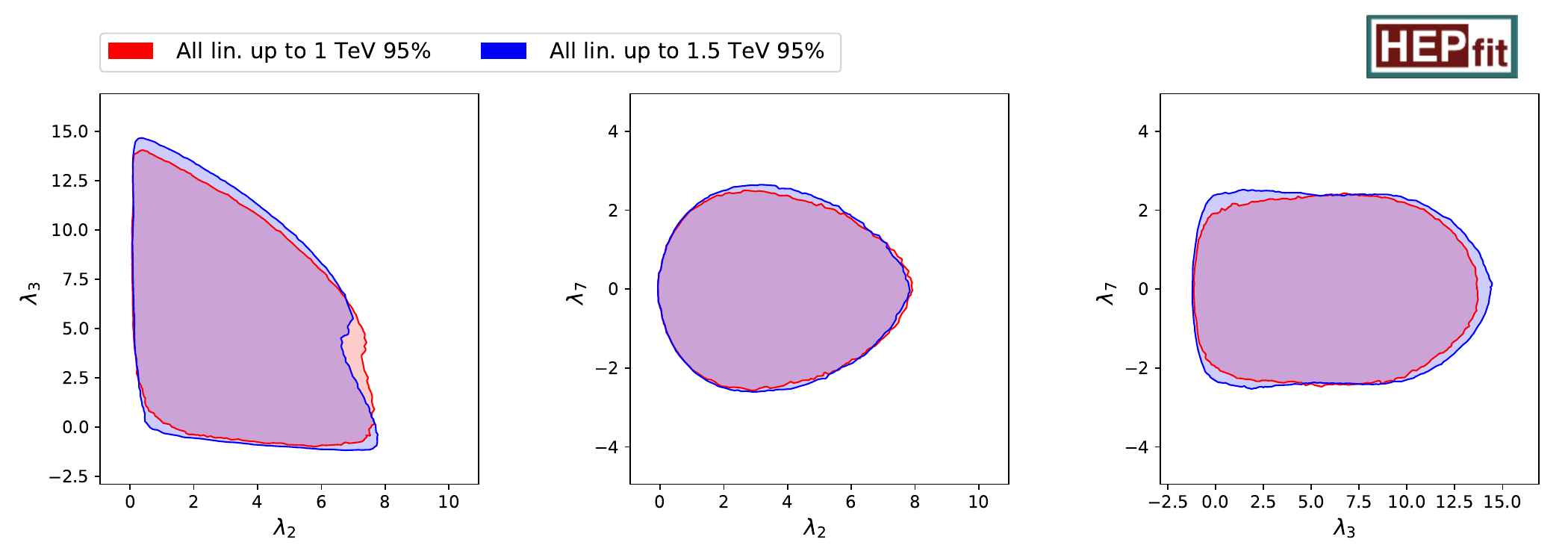}
	\vspace*{-0.3cm}
	\caption{Correlation among the quartic couplings.}
	\label{fig:Lambdas}
\end{figure}

\FloatBarrier

Finally, in Fig.~\ref{fig:Lambdas} we plot the correlations among the quartic couplings of the scalar potential. These couplings are mainly constrained by the theoretical bounds (perturbative unitarity, boundness from below and vacuum stability) and therefore their constraints do not depend on the mass range studied. Indeed, these limits are approximately the same as those of Fig.~\ref{fig:theo_lambdas},
in which only the theoretical considerations have been taken into account.

\subsection{CDF $W$ mass measurement}
\label{sec:CDF-MW}

\begin{figure}[h!]
	\centering
	\hspace*{-0.6cm}
	\includegraphics[scale=0.45]{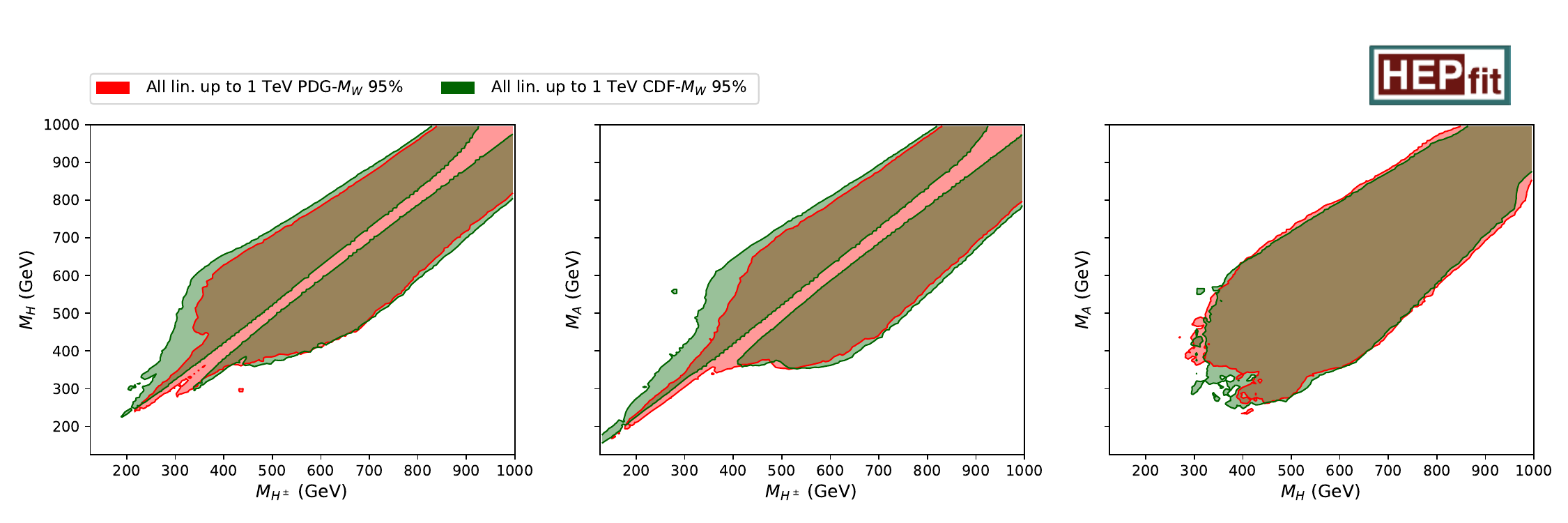}
	\vspace*{-3mm}
	\caption{Correlations among the masses of the new scalars, with and without including the CDF measurement of $M_W$.}
	\label{fig:All_Masses_CDF}
\end{figure}

\begin{figure}[h!]
	\centering
	\hspace*{-0.6cm}
	\includegraphics[scale=0.48]{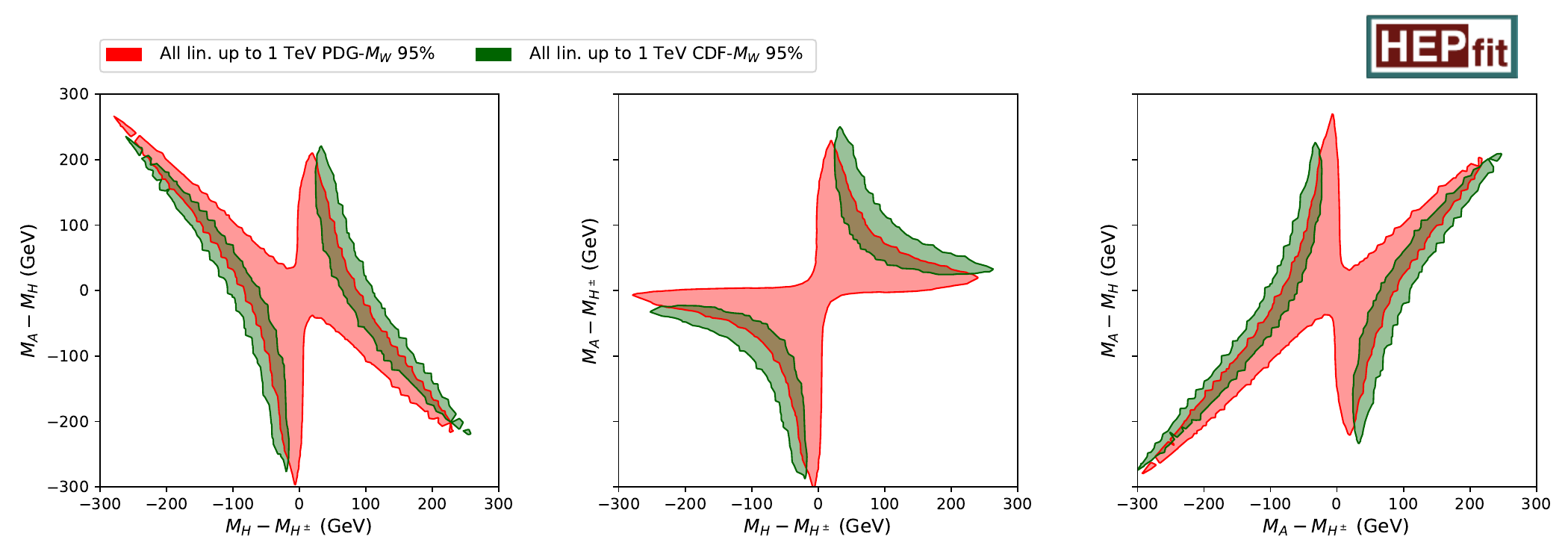}
	\vspace*{-3mm}
	\caption{Correlations among the scalar mass splittings, with and without including the CDF measurement of $M_W$.}
	\label{fig:Delta_mass_vs_delta_mass_CDF}
\end{figure}

As mentioned before, last year the CDF collaboration released a new measurement of the mass of the $W$ boson \cite{CDF:2022hxs} with a 7$\sigma$ tension with the SM prediction. Since there is still a 
controversy in the community, we decided not to include this last measurement in our main analysis but we provide here the results of a global fit including that information. In particular, we take the values of the oblique parameters ($S$ and $T$) from Ref.~\cite{deBlas:2022hdk}. Using that data as an input, we obtain a posterior value for $M_W=80.4178 \pm 0.0091$ GeV which is compatible with the CDF measurement ($M_W^{\mathrm{CDF}}=80.433 \pm 0.009$) within a 95\% probability. The additional scalars of the ATHDM would then be able to explain this result, if confirmed.\footnote{Of course we are not able to explain the 4$\sigma$ incompatibility between the ATLAS \cite{ATLAS:2017rzl,ATLAS:2023fsi} and CDF  \cite{CDF:2022hxs} measurements.} As can be seen in Figs.~\ref{fig:All_Masses_CDF} and \ref{fig:Delta_mass_vs_delta_mass_CDF}, in order to generate this additional contribution to $M_W$ we need a mass difference among the charged scalar and the neutral scalars of, at least, few tens of GeVs. Besides this mass difference, the results for the other parameters are very similar to the ones obtained in the baseline fit, as expected. 


\FloatBarrier

\section{Conclusion}
\label{sec:conclusion}

In this work we have performed an extensive phenomenological analysis of the CP-conserving Aligned Two-Higgs-Doublet model, using the \texttt{HEPfit} package. These results update the previous work of Ref.~\cite{Eberhardt:2020dat},\footnote{In this work we have adopted as baseline the same code used in Ref.~\cite{Eberhardt:2020dat}, but we have also included some relevant modifications. We thank here the developers of the previous version who are not signing this paper, O.~Eberhardt and A.~Pe\~nuelas.} although we use here a slightly different set of parameters. The observed SM Higgs boson at a mass of 125 GeV has been assumed to be the lightest scalar particle 
and we have, therefore, focused our study in heavy new-physics scalars.
More details on the region of the parameter space that we have scanned
can be found in Section~\ref{sec:set_up}.

We have incorporated the theoretical requirements of perturbative unitarity, boundedness of the scalar potential from below and vacuum stability,
and the most relevant experimental constraints, including direct searches at the LHC, Higgs signal strengths, electroweak precision observables and flavour data.
The main results of the global fit, combining all theoretical and experimental constraints at the same time,
are shown in Section \ref{sec:globa_fits_results}.

We have a total of ten new-physics parameters to fit. Three of them are the Yukawa alignment parameters for up-type quarks ($\varsigma_u$), down-type quarks ($\varsigma_d$) and leptons ($\varsigma_\nl$). The other seven come from the scalar potential, where we have chosen the three heavy scalar masses ($M_H$, $M_A$ and $M_{H^\pm}$), the mixing angle among the neutral scalars ($\tilde{\alpha}$), and three quartic couplings ($\lambda_2$, $\lambda_3$ and $\lambda_7$). The only relevant constraints on these quartic couplings come from theoretical considerations. The mixing angle is mainly constrained by the Higgs signal strengths and lower limits on the masses of the new scalars are obtained from direct searches and flavour observables (which constrain the mass of the charged scalar). The mass difference among the charged scalar and (at least one of) the neutral scalars is highly constrained by the oblique parameters. The theoretical requirements also provide relevant constraints on the mass splittings and on the mixing angle for high values of the charged scalar mass. Finally, the Yukawa alignment parameters are mainly constrained by the direct searches and flavour observables.

The marginalised probabilities for all the parameters can be found in Tab.~\ref{tab:marginalised_results}. In this table we show results for a fit in which the scan on the masses is done up to 1 TeV and one in which the scan is done up to 1.5 TeV. 
These ranges of masses have been chosen according to the ranges testable at the LHC.
Obviously, the results differ according to the range considered, since higher masses are always favoured. 
Note also that the lower bounds on the masses are highly correlated with the values of the Yukawa alignment parameters. If we set these parameters to sufficiently low values, the constraints on the masses vanish. For masses up to 1 TeV, values of $|\varsigma_u|>0.7$ , $|\varsigma_d|>12$,  $|\varsigma_\nl|>30$ and  $|\tilde{\alpha}|>0.06$ lay outside the 95\% probability region.

Besides performing this global fit with the state of the art data, we have also studied 
two additional measurements that, at the moment, deviate from the SM prediction: the muon $(g-2)$ \cite{Muong-2:2021ojo,Muong-2:2023cdq} and the new measurement of $M_W$ from the CDF collaboration \cite{CDF:2022hxs}. In both cases we have not included them in the global fit because they are still controversial, although for different reasons.
Nevertheless, we have shown that the ATHDM has enough flexibility to accommodate both measurements within its parameter space, while being still compatible with all other constraints.

\section*{Acknowledgement}

We thank the developers of the previous global fit to this model, Otto Eberhardt and Ana Pe\~nuelas.
We also thank Luca Silvestrini, Ayan Paul and Antonio Coutinho for some useful discussions and technical support regarding {\tt HEPfit}. 
This work has been supported in part by Generalitat Valenciana (Grant PROMETEO/2021/071) and by
MCIN/AEI/10.13039/501100011033 (Grant No. PID2020-114473GB-I00). The work of VM has been supported by the Italian Ministry of Research (MUR) under
the grant PRIN20172LNEEZ.

\appendix

\section{Data Included}

\label{sec:data_included}

The following tables compile all experimental references included in the fit for the direct searches and the Higgs signal strengths.


\vspace*{8mm}
\begin{table}[h!]
	{\small
		\begin{center}
			\begin{tabular}{|c |c| c c c| c c c|}
				\hline
				\multicolumn{1}{|c}{\textbf{Production}} & \textbf{Decay} & \textbf{Reference} & \textbf{$L$} & \textbf{$\sqrt{s}$} & \textbf{Reference} & \textbf{$L$} & \textbf{$\sqrt{s}$}\\[3pt]
				\multicolumn{1}{|c}{}& & & \textbf{[fb$^{-1}$]} & \textbf{[TeV]} & & \textbf{[fb$^{-1}$]} & \textbf{[TeV]} \\[3pt]
				\cline{3-8}
				
				\multicolumn{1}{|c}{}& &\multicolumn{3}{c|}{\textbf{ATLAS}}&\multicolumn{3}{c|}{\textbf{CMS}}\\
				\hline
				\multirow{ 15}{*}{\makecell{ ggF \\ \\ VBF \\ \\ Vh \\ \\ tth } }&  \multirow{ 1}{*}{$c \bar{c}$}  & \cite{ATLAS:2021zwx} & 139 & 13 & \cite{CMS:2019hve} & 35.9 & 13 \\
				\cline{2-8}

				& \multirow{ 2}{*}{$b \bar{b}$}  & \cite{ATLAS:2014vuz,ATLAS:2015utn} & 4.7/20.3 & 7/8 & \cite{CMS:2013poe,CMS:2014tll} & 5.1/18.9 & 7/8 \\
				&  & \cite{ATLAS:2020bhl,ATLAS:2020fcp,ParticleDataGroup:2022pth}  &  126/139  & 13 & \cite{CMS:2016mmc,CMS:2018nsn,ParticleDataGroup:2022pth}   & 2.3/ 41.3 & 13 \\\cline{2-8}

				& \multirow{ 2}{*}{$ \gamma\gamma$} & \multirow{ 1}{*}{\cite{ ATLAS:2014cnc}}   & \multirow{ 1}{*}{4.5/20.3}  & \multirow{ 1}{*}{7/8} & \multirow{ 1}{*}{\cite{CMS:2014afl}} & \multirow{ 1}{*}{5.1/19.7} & \multirow{ 1}{*}{7/8}\\
				&  & \multirow{ 1}{*}{\cite{ATLAS:2020pvn}}& \multirow{ 1}{*}{139} & \multirow{ 1}{*}{13}  & \multirow{ 1}{*}{ \cite{CMS:2021kom}} & \multirow{ 1}{*}{137} & \multirow{ 1}{*}{13} \\\cline{2-8}

				& \multirow{ 2}{*}{$\mu^+ \mu^-$}  & \cite{ ATLAS:2016neq} & 5/20& 7/8 & \cite{ ATLAS:2016neq} & 5/20& 7/8 \\
				&  & \cite{ATLAS:2020fzp}  &  139  & 13 & \cite{CMS:2020xwi}   & 137 & 13 \\\cline{2-8}
				&  \multirow{ 2}{*}{$\tau^+ \tau^-$ } & \cite{ATLAS:2015xst}& 4.5/20.3 & 7/8 & \cite{CMS:2014wdm}  & 4.9/19.7 & 7/8 \\
				&   &\cite{ATLAS:2020qdt} & 139 & 13&  \cite{CMS:2020gsy}  & 137 &13 \\\cline{2-8}
				&   \multirow{ 2}{*}{$WW$} & \cite{ ATLAS:2014aga, ATLAS:2015muc}  & 25, 4.5/20.3 & 7/8 & \cite{CMS:2013zmy} & 4.9/19.4& 7/8 \\
				&  & \cite{ATLAS:2020qdt} &  139  & 13 &  \cite{CMS:2020gsy} &  137 & 13 \\\cline{2-8}
				&  \multirow{ 2}{*}{$Z\gamma$ } & \cite{ATLAS:2015egz} & 4.7/20.3 & 7/8 &\cite{CMS:2013rmy}  & 5/19.6 & 7/8 \\
				&  & \cite{ATLAS:2020qcv} & 139 & 13 & \cite{CMS:2018myz}  & 35.9 & 13 \\\cline{2-8}
				&   \multirow{ 2}{*}{$ZZ$} & \cite{ATLAS:2014kct}&  4.5/20.3& 7/8 & \cite{CMS:2014fzn} & 5.1/19.7 & 7/8 \\
				&  & \cite{ATLAS:2020rej} & 139 & 13 &\cite{CMS:2020gsy} & 137 & 13 \\\hline
				
			\end{tabular}
			\caption{Higgs signal strengths measured by ATLAS and CMS. Note that not all the decays have been measured for all the production channels. }
			\label{Tab:HiggsStrengths}
		\end{center}
	}
\end{table}


\begin{table}[htb]
\begin{center}
\begin{tabular}{| l | l | l c | c | c |}
\hline
\textbf{Label} &\textbf{Channel} & \multicolumn{2}{| l |}{\textbf{Experiment}} & \textbf{Mass range} & ${\cal L}$ \\
&&&& \textbf{[TeV]} & \textbf{[fb$^{-1}$]} \\[1pt]
\hline
\cellcolor{color8TeV} $A_{8}^{\tau\nu}$ & $pp\to H^\pm \to \tau^\pm \nu $ & ATLAS &\cite{ATLAS:2014otc} & [0.18;1] & 19.5\\
\hline
\cellcolor{color8TeV} $C_{8}^{\tau\nu}$  & $pp\to H^{\RF \pm} \to \tau^{\RF \pm} \nu $ & CMS &\cite{CMS:2015lsf}& [0.18;0.6] & 19.7\\
\hline
  \cellcolor{color13TeV}  $A_{13}^{\tau\nu}$  & \multirow{2}{*}{$pp\to H^{\pm} \to \tau^\pm \nu $ }& \multirow{1}{*}{ATLAS} &  \cite{ATLAS:2018gfm} & [0.09;2] & 36.1\\

\cellcolor{color13TeV} $C_{13}^{\tau\nu}$ & & \multirow{1}{*}{CMS} &    \cite{CMS:2019bfg} & [0.08;3]  & 35.9 \\
\hline
\hline
\cellcolor{color8TeV} $A_{8}^{tb}$ & $pp\to H^\pm \to t b $ & ATLAS & \cite{ATLAS:2015nkq} & [0.2;0.6] & 20.3\\
\hline
\cellcolor{color8TeV} $C_{8}^{tb}$ & $pp\to H^{\RF \pm} \to t b $ & CMS & \cite{CMS:2015lsf} & [0.18;0.6] & 19.7\\
\hline
\cellcolor{color13TeV} $A_{13}^{tb}$ & \multirow{2}{*}{$pp\to H^\pm \to tb $} & \multirow{1}{*}{ATLAS} &   \cite{ATLAS:2021upq} & [0.2;2] & 139 \\\cline{3-6}
\cellcolor{color13TeV} $C_{13}^{tb}$ &  & CMS & \cite{CMS:2020imj} & [0.2;3] & 35.9\\
\hline
\end{tabular}
\caption{Direct searches for charged  scalars.}
\label{tab:directsearches4}
\end{center}
\end{table}

\begin{table}[htb]
\centering
{\renewcommand{\arraystretch}{1.0}
\resizebox{\textwidth}{!}{%
\begin{tabular}{| l | l | l c | c | c |}
\hline
\textbf{Label} &\textbf{Channel} & \multicolumn{2}{| l |}{\textbf{Experiment}} & \textbf{Mass range} & ${\cal L}$ \\
&&&& \textbf{[TeV]} & \textbf{[fb$^{-1}$]} \\[1pt]
\hline
\hline
\cellcolor{color8TeV} $A_{8}^{hh}$ &$gg\to \varphi_i^0 \to hh$ & ATLAS &\cite{ATLAS:2015sxd} & [0.26;1] & 20.3\\
\hline
\cellcolor{color8TeV} $C_{8}^{4b}$  &$pp\to  \varphi_i^0 \to hh \to (bb) (bb)$ & CMS &\cite{CMS:2015jal} & [0.27;1.1] & 17.9\\
\hline
\cellcolor{color8TeV} $C_{8}^{2\gamma2b}$  &$pp\to  \varphi_i^0 \to hh \to (bb) (\gamma \gamma)$ & CMS & \cite{CMS:2016cma} & [0.260;1.1] & 19.7\\
\hline
\cellcolor{color8TeV} $C_{8g}^{2b2\tau}$ &$gg\to  \varphi_i^0 \to hh \to (bb) (\tau\tau)$ & CMS & \cite{CMS:2015uzk} & [0.26;0.35] & 19.7\\
\hline
\cellcolor{color8TeV} $C_{8}^{2b2\tau}$  &$pp\to  \varphi_i^0 \to hh [\to (bb) (\tau\tau)]$ & CMS & \cite{CMS:2017yfv} & [0.35;1] & 18.3\\
\hline
\cellcolor{color13TeV} $A_{13}^{4b}$& \multirow{3}{*}{$pp \to  \varphi_i^0 \to hh \to (bb) (bb)$} & ATLAS & \cite{ATLAS:2022hwc} & [0.251;5] & 139\\
\cellcolor{color13TeV} $C_{13,1}^{4b}$  & &  CMS & \cite{CMS:2018qmt} & [0.26;1.2] & 35.9\\
\cellcolor{color13TeV} $C_{13,2}^{4b}$  & &  CMS & \cite{CMS:2021qvd} & [1;3] & 138\\
\hline
\cellcolor{color13TeV} $C_{13}^{4W}$& \multirow{1}{*}{$pp \to  \varphi_i^0 \to hh \to (WW)(WW)/(WW)(\tau\tau)/(\tau\tau)(\tau\tau)$} & CMS & \cite{CMS:2022kdx} & [0.25;1]  & 138\\
\hline
\cellcolor{color13TeV} $A_{13}^{2\gamma2b}$  & $pp \to  \varphi_i^0 \to hh [\to (bb) (\gamma \gamma)]$ & ATLAS &  \cite{ATLAS:2021ifb} & [0.251;1] & 139 \\
\cellcolor{color13TeV} $C_{13}^{2\gamma2b}$  & $pp \to  \varphi_i^0 \to hh \to (bb) (\gamma \gamma)$ & CMS & \cite{CMS:2018tla} & [0.25;0.9] & 35.9\\
\hline
\cellcolor{color13TeV} $A_{13,1}^{2b2\tau}$ & \multirow{2}{*}{$pp \to \varphi_i^0 \to hh \to (bb) (\tau \tau)$} & ATLAS &   \cite{ATLAS:2021fet} & [0.251;1.6] & 139\\
\cellcolor{color13TeV} $A_{13,1}^{2b2\tau}$ &  & ATLAS &  \cite{ATLAS:2020azv} & [1;3] & 139\\
\cellcolor{color13TeV} $C_{13,1}^{2b2\tau}$ & \multirow{2}{*}{$pp \to  \varphi_i^0 \to hh [\to (bb) (\tau \tau)]$} & CMS & \cite{CMS:2017hea} & [0.25;0.9] & 35.9\\
\cellcolor{color13TeV} $C_{13,2}^{2b2\tau}$ &  & CMS & \cite{CMS:2018kaz} & [0.9;4] & 35.9\\

\hline
\cellcolor{color13TeV} $C_{13}^{2b2V}$ & $pp \to  \varphi_i^0 \to hh \to (bb) (VV\to \ell \nu \ell \nu)$ & CMS & \cite{CMS:2017rpp} & [0.26;0.9] & 35.9\\
\hline
\cellcolor{color13TeV} $C_{13}^{2b2W}$ & $pp \to  \varphi_i^0 \to hh \to (bb) (WW\to q \bar{q} \ell \nu)$ & CMS & \cite{CMS:2019noi} & [0.8;3.5] & 35.9\\
\hline
\cellcolor{color13TeV} $C_{13}^{2b2Z}$ & $pp \to  \varphi_i^0 \to hh \to (bb) (ZZ \to \ell \ell j j)$ & CMS & \cite{CMS:2020jeo} & [0.26;1] & 35.9\\
\hline
\cellcolor{color13TeV} $C_{13}^{2b2Z}$ & $pp \to  \varphi_i^0 \to hh \to (bb) (ZZ \to \ell \ell \nu \nu)$ & CMS & \cite{CMS:2020jeo} & [0.26;1] & 35.9\\
\hline
\cellcolor{color13TeV} $A_{13}^{2b2W}$ & $pp \to  \varphi_i^0 \to hh [\to (bb) (WW)]$ & ATLAS &  \cite{ATLAS:2018fpd} & [0.5;3] & 36.1\\

\hline
\cellcolor{color13TeV} $C_{13}^{2b2W}$ & $pp \to  \varphi_i^0 \to hh [\to (bb) (WW/\tau\tau\to (q\bar{q}/\ell\nu) \ell\nu)]$ & CMS & \cite{CMS:2021roc} & [0.8;4.5] & 138\\
\hline
\cellcolor{color13TeV} $A_{13}^{2\gamma2W}$  & $gg \to  \varphi_i^0 \to hh \to (\gamma \gamma) (WW)$ & ATLAS & \cite{ATLAS:2018hqk} & [0.26;0.5] & 36.1\\ 
\hline
\hline
\cellcolor{color8TeV} $A_{8}^{bbZ}$ &$gg\to  \varphi_i^0 \to hZ \to (bb) Z$ & ATLAS & \cite{ATLAS:2015kpj} & [0.22;1] & 20.3\\
\hline
\cellcolor{color8TeV} $C_{8}^{2b2\ell}$ &$gg\to  \varphi_i^0 \to hZ \to (bb) (\ell \ell)$ & CMS &\cite{CMS:2015flt} & [0.225;0.6] &19.7\\
\hline
\cellcolor{color8TeV} $A_{8}^{\tau\tau Z}$ &$gg\to  \varphi_i^0 \to hZ \to (\tau\tau) Z$ & ATLAS & \cite{ATLAS:2015kpj} & [0.22;1] & 20.3\\
\hline
\cellcolor{color8TeV} $C_{8}^{2\tau2\ell}$  &$gg\to  \varphi_i^0 \to hZ \to (\tau\tau) (\ell \ell)$ & CMS & \cite{CMS:2015uzk} & [0.22;0.35] & 19.7\\
\hline
\cellcolor{color13TeV} $A_{13}^{bbZ}$  & \multirow{5}{*}{$gg\to  \varphi_i^0 \to hZ \to (bb) Z$} & ATLAS & \cite{ATLAS:2017xel} & [0.2;2] & 36.1\\
\cellcolor{color13TeV} $C_{13,1}^{bbZ}$  & & CMS & \cite{CMS:2019qcx} & [0.22;0.8] & 35.9\\
\cellcolor{color13TeV} $C_{13,2}^{bbZ}$  & & CMS &  \cite{CMS:2018ljc} & [0.8;2] & 35.9\\
\hline
\cellcolor{color13TeV} $C_{13,3}^{bbZ}$  & $gg\to  \varphi_i^0 \to (h \to \tau\tau) (Z \to \ell \ell)$ & CMS & \cite{CMS:2019kca} & [0.22;0.4] & 35.9\\
\hline
\cellcolor{color13TeV} $A_{13b}^{bbZ}$  & \multirow{3}{*}{$bb\to  \varphi_i^0 \to hZ \to (bb) Z$} & ATLAS & \cite{ATLAS:2017xel} & [0.2;2] & 36.1\\
\cellcolor{color13TeV} $C_{13b,1}^{bbZ}$  & & CMS & \cite{CMS:2019qcx} & [0.22;0.8] & 35.9\\
\cellcolor{color13TeV} $C_{13b,2}^{bbZ}$  & & CMS & \cite{CMS:2018ljc} & [0.8;2] & 35.9\\
\hline
\hline
\cellcolor{color8TeV} $C_{8, 1}^{\varphi_2^0 Z}$& $pp\to \varphi_3^0  \to \varphi_2^0 Z \to (bb) (\ell\ell)$ & CMS & \cite{CMS:2016xnc} & [0.04;1] & 19.8\\
\hline
\cellcolor{color8TeV} $C_{8, 2}^{\varphi_2^0 Z}$& $pp\to \varphi_3^0 \to \varphi_2^0 Z \to (\tau \tau) (\ell\ell)$ & CMS & \cite{CMS:2016xnc} & [0.05;1] & 19.8\\
\hline
\cellcolor{color13TeV} $A_{13}^{\varphi^0 Z}$& $gg\to \varphi_{3}^0 \to \varphi_2^0 Z \to (bb) Z$ & ATLAS &  \cite{ATLAS:2020gxx} & [0.13/0.23;0.7/0.8]  & 139\\
\hline
\cellcolor{color13TeV} $A_{13b}^{\varphi^0 Z}$& $bb\to \varphi_{3}^0 \to \varphi_2^0 Z \to (bb) Z$ & ATLAS &  \cite{ATLAS:2020gxx} & [0.13/0.23;0.7/0.8]  & 139\\
\hline
\end{tabular}
}
}
\caption{Direct searches for neutral heavy scalars, $\varphi_i^0 = H, A$, with final states including the SM Higgs boson or other neutral scalars. $\varphi_3$ denotes the heaviest scalar,  $V = W,Z$, $\ell = e, \mu$. The parenthesis show the final decay of the SM particles produced from the NP particles. The square brackets are used when the values of $\sigma\cdot \mathcal{B}$ are shown in terms of the primary decay (i.e. the NP particle decay) but a particular decay channel of the SM particles is used to obtain those values. See Section~\ref{sec:direct_searches} for more details.
}
\label{tab:directsearches3}
\end{table}

\begin{table}[htb]
	\centering
	{\renewcommand{\arraystretch}{0.8}
		\resizebox{0.88\textwidth}{!}{%
			\begin{tabular}{|l|l|lc|c|c|}
				\hline
				\textbf{Label} &\textbf{Channel} & \multicolumn{2}{| l |}{\textbf{Experiment}} & \textbf{Mass range} & ${\cal L}$ \\
				&&&& \textbf{[TeV]} & \textbf{[fb$^{-1}$]} \\[1pt]
				\hline
				\hline

				\cellcolor{color13TeV} $A_{13b}^{tt}$  & $bb \to \varphi_i^0 \to tt$ & ATLAS & \cite{ATLAS:2016btu} & [0.4;1] & 13.2\\
				\hline
				\hline
				
				\cellcolor{color13TeV}$C_{13t}^{tt}$  & $tt/tW/tq \to \varphi_i^0 \to tt$ & CMS & \cite{CMS:2019rvj} & [0.35;0.65] & 137\\
				\hline
				\cellcolor{color13TeV}$A_{13t}^{tt}$  & $tt \to \varphi_i^0 \to tt$ & ATLAS & \cite{ATLAS:2022ohr} & [0.4;1] & 137\\
				\hline
				\hline

				\cellcolor{color8TeV} $C_{8b}^{bb}$ &$bb \to \varphi_i^0 \to bb$ & CMS & \cite{CMS:2015grx} & [0.1;0.9] & 19.7\\
				\hline
				\cellcolor{color8TeV} $C_{8}^{bb}$  &$gg \to \varphi_i^0\to bb$ & CMS & \cite{CMS:2018kcg} & [0.33;1.2] & 19.7\\
				\hline
				\cellcolor{color13TeV} $C_{13}^{bb}$ & $pp \to \varphi_i^0\to bb$ & CMS & \cite{CMS:2016ncz} & [0.55;1.2] & 2.69\\
				\hline
				\cellcolor{color13TeV} $C_{13b}^{bb}$ & $bb \to \varphi_i^0\to bb$ & CMS & \cite{CMS:2018hir} & [0.3;1.3] & 35.7\\
				\hline\hline

				\cellcolor{color13TeV}$A_{13}^{bb}$ & $pp \to \varphi_i^0\to bb$  ($\geq 1$ b-jet)& ATLAS & \cite{ATLAS:2018tfk} & [1.4;6.6] & 36.1\\
				\hline
				
				\cellcolor{color13TeV}$A_{13}^{bb}$ & $pp \to \varphi_i^0\to bb$ & ATLAS & \cite{ATLAS:2018tfk} & [0.6;1.25] & 24.3\\
				\hline
				
				\cellcolor{color13TeV}$A_{13}^{bb}$ & $pp \to \varphi_i^0\to bb$ & ATLAS & \cite{ATLAS:2018tfk} & [1.25;6.2] & 36.1\\
				\hline

				\cellcolor{color13TeV}$A_{13b}^{bb}$ & $bb \to \varphi_i^0\to bb$ & ATLAS & \cite{ATLAS:2019tpq} & [0.45;1.4] & 27.8\\
				\hline
				\hline
				
				\cellcolor{color13TeV}$C_{13}^{bb}$ & $pp \to \varphi_i^0\to bb$ & CMS & \cite{CMS:2018pwl} & [0.05;0.35] & 35.9\\
				\hline
				\hline

				\cellcolor{color8TeV} $C_{8b}^{\mu\mu}$ &$bb\to \varphi_i^0 \to \mu\mu$ & CMS &\cite{CMS:2015ooa} & [0.12;0.5] & 19.3 \\
				\hline
				\cellcolor{color8TeV} $C_{8}^{\mu\mu}$ &$gg\to \varphi_i^0 \to \mu\mu$ & CMS &\cite{CMS:2015ooa} & [0.12;0.5] & 19.3 \\
				\hline
				
				\cellcolor{color13TeV} $C_{13b}^{\mu\mu}$ &$bb\to \varphi_i^0 \to \mu\mu$ & CMS &\cite{CMS:2019mij} & [0.14;1] & 35.9 \\
				\hline
				\cellcolor{color13TeV} $C_{13}^{\mu\mu}$ &$gg\to \varphi_i^0 \to \mu\mu$ & CMS &\cite{CMS:2019mij} & [0.14;1] & 35.9 \\
				\hline
				
				\cellcolor{color13TeV} $A_{13b}^{\mu\mu}$ &$bb\to \varphi_i^0 \to \mu\mu$ & ATLAS &\cite{ATLAS:2019odt} & [0.2;1] & 36.1 \\
				\hline
				\cellcolor{color13TeV} $A_{13}^{\mu\mu}$ &$gg\to \varphi_i^0 \to \mu\mu$ & ATLAS &\cite{ATLAS:2019odt} & [0.2;1] & 36.1 \\
				\hline\hline

				\cellcolor{color8TeV} $A_{8}^{\tau\tau}$ &\multirow{2}{*}{$gg\to \varphi_i^0 \to \tau\tau$} & ATLAS &\cite{ATLAS:2014vhc} & [0.09;1] & 20 \\
				\cellcolor{color8TeV} $C_{8}^{\tau\tau}$  & & CMS &\cite{CMS:2015mca} &  [0.09;1]  &19.7 \\
				\hline
				\cellcolor{color8TeV} $A_{8b}^{\tau\tau}$  &\multirow{2}{*}{$bb\to \varphi_i^0 \to \tau\tau$} & ATLAS &\cite{ATLAS:2014vhc} & [0.09;1] & 20 \\
				\cellcolor{color8TeV} $C_{8b}^{\tau\tau}$ & & CMS & \cite{CMS:2015mca}& [0.09;1] & 19.7 \\
				\hline
				
				\cellcolor{color13TeV} $A_{13}^{\tau\tau}$ &$gg \to \varphi_i^0\to \tau \tau$ & ATLAS & \cite{ATLAS:2016ivh} & [0.2;1.2] & 3.2\\
				\cellcolor{color13TeV} $A_{13b}^{\tau\tau}$ &$bb \to \varphi_i^0\to \tau \tau$ & ATLAS & \cite{ATLAS:2016ivh} & [0.2;1.2] & 3.2\\
				\hline\hline

				\cellcolor{color13TeV} $A_{13}^{\tau\tau}$ &$gg \to \varphi_i^0\to \tau \tau$ & ATLAS & \cite{ATLAS:2020zms} & [0.2;2.5] & 139\\
				\cellcolor{color13TeV} $A_{13b}^{\tau\tau}$ &$bb \to \varphi_i^0\to \tau \tau$ & ATLAS & \cite{ATLAS:2020zms} & [0.2;2.5] & 139\\
				\hline\hline
				
				\cellcolor{color13TeV} $C_{13}^{\tau\tau}$ &$gg \to \varphi_i^0\to \tau \tau$ & CMS & \cite{CMS:2022rbd} & [0.06;3.5] & 138\\
				\cellcolor{color13TeV} $C_{13b}^{\tau\tau}$ &$bb \to \varphi_i^0\to \tau \tau$ & CMS & \cite{CMS:2022rbd} & [0.06;3.5] & 138\\
				\hline\hline
				
				\cellcolor{color13TeV} $A_{13}^{\tau\tau}$ &\multirow{2}{*}{$gg \to \varphi_i^0\to \tau \tau$} & ATLAS & \cite{ATLAS:2017eiz} & [0.2;2.25] & 36.1\\[-1pt]
				\cellcolor{color13TeV} $C_{13}^{\tau\tau}$ &  & CMS & \cite{CMS:2018rmh} & [0.09;3.2] & 35.9\\
				\hline
				\cellcolor{color13TeV} $A_{13b}^{\tau\tau}$ &\multirow{2}{*}{$bb \to \varphi_i^0\to \tau \tau$} & ATLAS & \cite{ATLAS:2017eiz} & [0.2;2.25] & 36.1\\[-1pt]
				\cellcolor{color13TeV} $C_{13b}^{\tau\tau}$ & & CMS & \cite{CMS:2018rmh} & [0.09;3.2] & 35.9\\
				\hline
				\hline
				\cellcolor{color8TeV} $A_{8}^{\gamma\gamma}$&$gg\to \varphi_i^0 \to \gamma\gamma$ & ATLAS &\cite{ATLAS:2014jdv} & [0.065;0.6] & 20.3 \\
				\hline
				\cellcolor{color13TeV} $C_{13}^{\gamma\gamma}$ & $gg \to \varphi_i^0\to \gamma \gamma$ & CMS & \cite{CMS:2016kgr} & [0.5;4] & 35.9\\ 
				\hline
				
				\cellcolor{color13TeV} $C_{13}^{\gamma\gamma}$ & $gg \to \varphi_i^0\to \gamma \gamma$ & CMS & \cite{CMS:2018dqv} & [0.5;5] & 35.9\\ 
				\cellcolor{color13TeV} $A_{13}^{\gamma\gamma}$ & $pp \to \varphi_i^0\to \gamma \gamma$ & ATLAS & \cite{ATLAS:2021uiz} & [0.15;3] & 139\\ 
				\hline
				
				\hline

				\cellcolor{color8TeV} $A_{8}^{Z\gamma}$  &\multirow{2}{*}{$pp\to \varphi_i^0 \to Z\gamma \to (\ell \ell) \gamma$} & ATLAS & \cite{ATLAS:2014lfk} & [0.2;1.6] & 20.3 \\
				\cellcolor{color8TeV} $C_{8}^{Z\gamma}$ & & CMS & \cite{CMS:2016all} & [0.2;1.2] & 19.7 \\
				\hline

				\cellcolor{color13TeV} $C_{13}^{\ell\ell\gamma}$  & $pp \to \varphi_i^0\to Z \gamma [\to (\ell \ell) \gamma ]$ & CMS & \cite{CMS:2017dyb} & [0.35;4] & 35.9\\
				\cellcolor{color13TeV} $C_{13}^{qq\gamma}$  & $pp \to \varphi_i^0\to Z \gamma [\to (qq) \gamma ]$ & CMS & \cite{CMS:2017dyb} & [0.35;4] & 35.9\\
				\cellcolor{color13TeV} $C_{13}^{Z\gamma}$  & $pp \to \varphi_i^0\to Z \gamma [\to (\ell \ell \,\&\, qq) \gamma ]$ & CMS & \cite{CMS:2017dyb} & [0.35;4] & 35.9\\
				\hline

				\cellcolor{color13TeV} $A_{13}^{\ell\ell\gamma}$  & $gg \to \varphi_i^0\to Z \gamma [\to (\ell \ell) \gamma ]$ & ATLAS & \cite{ATLAS:2017zdf} & [0.25;2.4] & 36.1\\
				\hline
				\cellcolor{color13TeV} $A_{13}^{qq\gamma}$  & $gg \to \varphi_i^0\to Z \gamma [\to (qq) \gamma ]$ & ATLAS & \cite{ATLAS:2018sxj} & [1;6.8] & 36.1\\
				\hline
				\cellcolor{color13TeV} $C_{8+13}^{Z\gamma}$  & $gg \to \varphi_i^0\to Z \gamma$ & CMS & \cite{CMS:2017dyb} & [0.35;4] & 35.9\\
				\hline
			\end{tabular}
		}
	}
	\caption{Direct searches for neutral heavy scalars, $\varphi_i^0 = H, A$, with quarks, leptons ($\ell= e,\mu$), photons and $Z\gamma$ final states. The parenthesis show the final decay of the SM particles produced from the NP particles. The square brackets are used when the values of $\sigma\cdot \mathcal{B}$ are shown in terms of the primary decay (i.e. the NP particle decay) but a particular decay channel of the SM particles is used to obtain those values. See Section~\ref{sec:direct_searches} for more details.}
	\label{tab:directsearches1}
\end{table}

\begin{table}[htb]
	\centering
	{\renewcommand{\arraystretch}{1.}
		\resizebox{\textwidth}{!}{%
			\begin{tabular}{|l|l|lc|c|c|}
				\hline
				\textbf{Label} &\textbf{Channel} & \multicolumn{2}{ l| }{\textbf{Experiment}} & \textbf{Mass range} & ${\cal L}$ \\
				&&&& \textbf{[TeV]} & \textbf{[fb$^{-1}$]} \\[1pt]
				\hline
				\cellcolor{color8TeV} $A_{8}^{ZZ}$  &$gg\to \varphi_i^0\to ZZ$ & ATLAS & \cite{ATLAS:2015pre}& [0.14;1] & 20.3 \\
				\hline
				\cellcolor{color8TeV} $A_{8V}^{ZZ}$  &$VV \to \varphi_i^0\to ZZ$ & ATLAS & \cite{ATLAS:2015pre}& [0.14;1] & 20.3 \\
				\hline
				\cellcolor{color13TeV} $A_{13}^{2\ell2L}$  & $gg\to \varphi_i^0 \to ZZ [\to (\ell \ell) (\ell \ell, \nu \nu)]$ & ATLAS & \cite{ATLAS:2017tlw} & [0.2;1.2] & 36.1\\
				\hline
				\cellcolor{color13TeV} $A_{13V}^{2\ell2L}$  & $VV\to \varphi_i^0\to ZZ [\to (\ell \ell) (\ell \ell, \nu \nu)]$ & ATLAS & \cite{ATLAS:2017tlw} & [0.2;1.2] & 36.1\\
				\hline

				\cellcolor{color13TeV} $A_{13}^{2\ell2L}$  & $gg\to \varphi_i^0 \to ZZ [\to (\ell \ell) (\ell \ell, \nu \nu)]$ & ATLAS & \cite{ATLAS:2020tlo} & [0.2;2] & 139\\
				\hline
				\cellcolor{color13TeV} $A_{13V}^{2\ell2L}$  & $VV\to \varphi_i^0\to ZZ [\to (\ell \ell) (\ell \ell, \nu \nu)]$ & ATLAS & \cite{ATLAS:2020tlo} & [0.2;2] & 139\\
				\hline

				\cellcolor{color13TeV} $A_{13}^{2L2q}$ & $gg\to \varphi_i^0\to ZZ [\to (\ell \ell, \nu \nu) (qq)]$ & ATLAS & \cite{ATLAS:2017otj} & [0.3;3] & 36.1\\
				\hline
				\cellcolor{color13TeV} $A_{13V}^{2L2q}$ & $VV\to \varphi_i^0\to ZZ [\to (\ell \ell, \nu \nu) (qq)]$ & ATLAS & \cite{ATLAS:2017otj} & [0.3;3] & 36.1\\
				\hline
				\cellcolor{color13TeV} $C_{13}^{2\ell2X}$ & $pp\to \varphi_i^0\to ZZ [\to (\ell \ell) (qq,\nu\nu,\ell\ell)]$ & CMS & \cite{CMS:2018amk} & [0.13;3] & 35.9\\
				\hline
				\cellcolor{color13TeV} $C_{13}^{2q2\nu}$ & $pp\to \varphi_i^0\to ZZ [\to (qq)(\nu\nu)]$ & CMS & \cite{CMS:2018ygj} & [1;4] & 35.9\\
				\hline
				\hline
				\cellcolor{color8TeV} $A_{8}^{WW}$  &$gg\to \varphi_i^0\to WW$ & ATLAS &\cite{ATLAS:2015iie}& [0.3;1.5] & 20.3 \\			
				\hline
				\cellcolor{color8TeV} $A_{8V}^{WW}$ &$VV \to \varphi_i^0\to WW$ & ATLAS & \cite{ATLAS:2015iie}& [0.3;1.5] & 20.3 \\
				
				\hline
				\hline
				\cellcolor{color13TeV} $C_{13V}^{WW}$ &$VV \to \varphi_i^0\to WW$ & CMS & \cite{CMS:2019bnu}& [0.2;3] & 35.9 \\
				\hline
				\hline
				
				\cellcolor{color13TeV} $A_{13}^{2(\ell\nu)}$  & $gg\to \varphi_i^0\to WW [\to (e \nu) (\mu \nu)]$ & ATLAS & \cite{ATLAS:2017uhp} & [0.2;4] & 36.1\\
				\hline
				\cellcolor{color13TeV} $A_{13V}^{2(\ell\nu)}$  & $VV\to \varphi_i^0\to WW [\to (e \nu) (\mu \nu)]$ & ATLAS & \cite{ATLAS:2017uhp} & [0.2;3] & 36.1\\
				\hline
				\cellcolor{color13TeV} $C_{13}^{2(\ell\nu)}$ & $(gg\!+\!VV)\to \varphi_i^0\to WW \to (\ell \nu) (\ell \nu)$ & CMS & \cite{CMS:2016jpd} & [0.2;1] & 2.3\\
				\hline
				\cellcolor{color13TeV} $A_{13}^{\ell\nu2q}$ & $gg\to \varphi_i^0\to WW[\to (\ell \nu) (qq)]$ & ATLAS & \cite{ATLAS:2017jag} & [0.3;3] & 36.1\\
				\hline
				\cellcolor{color13TeV} $A_{13V}^{\ell\nu2q}$ & $VV\to \varphi_i^0\to WW[\to (\ell \nu) (qq)]$ & ATLAS & \cite{ATLAS:2017jag} & [0.3;3] & 36.1\\
				\hline
				\cellcolor{color13TeV} $C_{13}^{\ell\nu2q}$ & $pp\to \varphi_i^0\to WW[\to (\ell \nu) (qq)]$ & CMS & \cite{CMS:2018dff} & [1;4.4] & 35.9\\
				\hline
				\hline
				\cellcolor{color8TeV} $C_{8}^{VV}$  & $pp \to \varphi_i^0\to VV$ & CMS & \cite{CMS:2015hra} & [0.145;1] & 24.8 \\
				\cellcolor{color13TeV} $A_{13}^{4q}$  & $pp\to \varphi_i^0\to VV [\to (qq) (qq)]$ & ATLAS & \cite{ATLAS:2017zuf} & [1.2;3] & 36.7\\
				\hline
				
			\cellcolor{color13TeV} $A_{13}^{VV}$  & $pp \to \varphi_i^0\to VV$ & ATLAS & \cite{ATLAS:2018sbw} & [0.3;3] & 36.1 \\	
			
			\cellcolor{color13TeV} $A_{13V}^{VV}$  & $VV \to \varphi_i^0\to VV$ & ATLAS & \cite{ATLAS:2018sbw} & [0.3;3] & 36.1 \\	
			\hline\hline
			
			\cellcolor{color13TeV} $A_{13}^{VV}$  & $gg \to \varphi_i^0\to VV$ & ATLAS & \cite{ATLAS:2020fry} & [0.2;5.2] & 139 \\	
			
			\cellcolor{color13TeV} $A_{13V}^{VV}$  & $VV \to \varphi_i^0\to VV$ & ATLAS & \cite{ATLAS:2020fry} & [0.2;5.2] & 139 \\	
			\hline
			\hline
			
			\cellcolor{color13TeV} $C_{13}^{WW}$  & $gg \to \varphi_i^0\to WW$ & CMS & \cite{CMS:2021klu} & [1,4.5] & 137 \\	
			
			\cellcolor{color13TeV} $C_{13V}^{WW}$  & $VV \to \varphi_i^0\to WW$ & CMS & \cite{CMS:2021klu} & [1;4.5] & 137 \\	
			\hline
			\end{tabular}
		}
	}
	\caption{Direct searches for neutral heavy scalars, $\varphi_i^0 = H, A$, with vector-boson final states. $V = W,Z$, $\ell = e, \mu$. The parenthesis show the final decay of the SM particles produced from the NP particles. The square brackets are used when the values of $\sigma\cdot \mathcal{B}$ are shown in terms of the primary decay (i.e. the NP particle decay) but a particular decay channel of the SM particles is used to obtain those values. See Section~\ref{sec:direct_searches} for more details.}
	\label{tab:directsearches2}
\end{table}

\FloatBarrier


\section{Non-contaminated Standard Model inputs}

In this work we have included as inputs the entries of the CKM matrix, in the Wolfenstein parametrisation \cite{Wolfenstein:1983yz}, and the oblique parameters \cite{Peskin:1990zt, Peskin:1991sw}. 
The experimental values of these parameters quoted by the Particle Data Group \cite{ParticleDataGroup:2022pth}  have been obtained through a SM fit of multiple observables, neglecting any new-physics contributions. Unfortunately, this assumption is not satisfied in our model for all the observables included in the fits, and we need to repeat those fits removing the problematic observables. 

\subsection{Non-contaminated CKM fit}
\label{sec:CKM_fit}

We extract the Wolfenstein parameters from the measured values of the CKM matrix elements (or ratios among them). 

\subsubsection{Determination of $|V_{ud}|$}

For $|V_{ud}|$ we use directly the value quoted by the PDG \cite{ParticleDataGroup:2022pth},
\begin{equation}
    |V_{ud}|=0.97373\pm 0.00031,
\end{equation}
since this value comes from superallowed $0^+\rightarrow 0^+$ nuclear $\beta$ decays \cite{Hardy:2020qwl} with completely negligible contamination from our additional scalars.

\subsubsection{Determination of $|V_{us}|$}

The determination of $|V_{us}|$ in the PDG \cite{ParticleDataGroup:2022pth} is based on semileptonic decays of kaons, averaging the electronic and muonic channel, and, independently, on the pure leptonic decay of kaons and pions to muons which provide the ratio $|V_{us}/V_{ud}|$. 
The scalar contribution to the semileptonic decays is highly suppressed by the light lepton masses. Therefore, as long as the mass of the kaons is much higher than the mass of the decaying lepton we can safely neglect the NP contribution in the semileptonic decays. This assumption holds for the semileptonic decay to electrons but it is not fulfilled for the muonic channel.  
 Therefore, for the semileptonic decays we only consider the decays to electrons, including the $K_L$, $K_S$ and $K^+$ decays, which give the average value 
 $|V_{us}f_+^K(0)|=0.21626\pm 0.00040$ \cite{Seng:2021nar}.
 Taking the form factor average.
$f_+(0)=0.9698\pm 0.0017$,  from $N_f= 2 + 1 + 1$ lattice QCD calculations \cite{FlavourLatticeAveragingGroupFLAG:2021npn}, gives $|V_{us}|=0.2230\pm 0.0006$.

As mentioned above, the alternative determination of $|V_{us}|$ is based on the leptonic kaon  decay. In this case, the SM contribution is helicity suppressed, making the scalar contribution relatively much higher. In general, the charged-scalar contribution to the leptonic decay of a pseudoscalar-meson  takes the form \cite{Jung:2010ik}
\begin{equation}
    \Gamma(P^+_{ij}\rightarrow l^+ \RF{\nu}_l) = \Gamma^{\rm SM}(P^+_{ij}\rightarrow l^+ \RF{\nu}_l)\; |1-\Delta_{ij}|^2\, ,
\end{equation}
with $i,j$ the flavour indices corresponding to the valence quarks of the meson, the new-physics correction
\begin{equation}
    \Delta_{ij}=\left(\frac{m_{P^+_{ij}}}{M_{H^\pm}}\right)^2 \varsigma_l^*\;\frac{\varsigma_u\, m_{ui}+\varsigma_d\, m_{dj}}{m_{ui}+ m_{dj}}\, ,
\end{equation}
and the SM contribution being related with the CKM matrix element by
\begin{equation}
    \Gamma^{\rm SM}(P^+_{ij}\rightarrow l^+ \RF{\nu}_l) = G_F^2\, m_l^2\, f^2_P\,|V_{ij}|^2 \, \frac{m_{P^+_{ij}}}{8\pi}\left(1-\frac{m_l^2}{m_{P^+_{ij}}}\right)^2
    \left(1+\delta_{\rm em}^{M\ell 2}\right)\, ,
\end{equation}
where $f_P$ is the meson decay constant and $\delta_{\rm em}^{M\ell 2}$ the electromagnetic radiative corrections.

In particular, in order to determine $|V_{us}|$, the ratio among the kaon and pion decay widths into muons is used, which gets a scalar contribution that is dominated by $2\Delta_{us}\approx 2\varsigma^*_l \varsigma_d m_K^2/M_{H^\pm}$. Since $\varsigma^*_l$ and $\varsigma_d$ can, in general, reach quite high values, we cannot neglect the scalar contribution in this case and, therefore, we cannot utilize this for the $|V_{us}|$ determination.


Note that the determination of $|V_{us}|$ obtained from the average of measurements involving leptonic decays ($|V_{us}|= 0.2252\pm 0.0005$) is not compatible by more than two standard deviations with the determination derived from the average of measurements involving semileptonic decays ($|V_{us}|= 0.2231\pm 0.0006$). When these two determinations are combined in the PDG average, the total uncertainty of the mean value is multiplied by a factor of two ($|V_{us}|= 0.2243\pm 0.0008$) in order to account for the discrepancy among them.
In our case, we could attribute the discrepancy to the new-physics effects. However, using only the determination from semileptonic decays generates a deviation from unitarity in the first row of the CKM matrix above the three-sigma level, when combined with $|V_{ud}|$ from nuclear $\beta$ decays. This `Cabibbo anomaly' is already present in the PDG  average, although at a slightly lower level because the leptonic kaon decay pushes
the central value of $|V_{us}|$ to a more favorable direction and the uncertainty is increased. In order to relax the deviation from unitarity below the three-sigma level, and also motivated by the PDG procedure, we follow a more conservative approach increasing also the uncertainty on $|V_{us}|$ by a factor of two (as well as the PDG does). The final value used in our fits is then 
\begin{equation}
    |V_{us}|=0.2230\pm 0.0012\, .
\end{equation}

\subsubsection{Determination of $|V_{cd}|$}

The PDG \cite{ParticleDataGroup:2022pth} average is obtained from semileptonic decays of $D$ mesons to light leptons, leptonic $D$ decays to muons and taus, and from neutrino scattering data.
The scalar contribution to the leptonic decay can be sizable and, therefore, we will only use the data coming from semileptonic decays
($|V_{cd}|=0.2330\pm0.0136$)
and the neutrino scattering data
($|V_{cd}|=0.230\pm0.011$). 
Averaging these two values we obtain
\begin{equation}
    |V_{cd}|=0.231\pm0.009\, .
\end{equation}
Note, however, that the uncertainty is significantly higher than the one of $|V_{us}|$, so the impact of this measurement on the CKM-fit is basically negligible.

\subsubsection{Determination of $|V_{cs}|$}

Similarly to the previous case, the determination of $|V_{cs}|$ is obtained from measurements of semileptonic decays of $D$ mesons and the leptonic decay of $D_s$, provided that the form factors are obtained from lattice QCD computations. We have dismissed the determination from leptonic decays and we have used only the one coming from semileptonic decays:
\begin{equation}
    |V_{cs}|=0.972\pm 0.007\, .
\end{equation}
As happens for $|V_{cd}|$, we have a much higher uncertainty compared to the light-quark data ($|V_{ud}|$) so, again, this observable could be neglected from our fit leaving the results unchanged. 

\subsubsection{Determinations of $|V_{cb}|$ and $|V_{ub}|$}

The methods used to determine $|V_{cb}|$ and $|V_{ub}|$ in the PDG do not receive sizable contributions from our additional scalars. Note that, as before, the leptonic decay of the $B$ mesons would be affected by the NP but these processes are not used for the current world average due to their \tp{large} uncertainty. The values quoted by the PDG \cite{ParticleDataGroup:2022pth} are
\begin{equation}
   |V_{cb}|=(40.8 \pm 1.4)\times 10^{-3},
\end{equation}
and
\begin{equation}
   |V_{ub}|=(3.82 \pm 0.20)\times 10^{-3}.
\end{equation}

\subsubsection{Determination of $|V_{td}/V_{ts}|$}

Finally, we will use the determination of $|V_{td}/V_{ts}|$, which is obtained from measurements of $B^0_{(s)}-\bar{B}^0_{(s)}$ meson mixing. Obviously, the additional scalars will contribute to this mixing but, once the ratio of the $B_d$ and $B_s$ transitions is taken, the new-physics contribution is highly suppressed since it is only present through
SU(3)-breaking effects. We adopt the value for $|V_{td}/V_{ts}|$ quoted by the PDG \cite{ParticleDataGroup:2022pth}:
\begin{equation}
   |V_{td}/V_{ts}|=(0.207 \pm 0.001 \pm 0.003).
\end{equation}

\subsubsection{CKM fit result}
Using as inputs all the measurements mentioned in the previous sections, we obtain the values of Tab.~\ref{tab:WolfResult} for the Wolfenstein parameters.

\begin{table}[h!]
\centering
\begin{tabular}{c|c|c|c|c|c}
    \multirow{2}{*}{ } & \multirow{2}{*}{Value} & \multicolumn{4}{c}{Correlation}  \\\cline{3-6}
    & & $ \lambda $ & $A$ & $\overline{\rho}$ & $ \overline{\eta}$\\ \hline
    $\lambda$ & $0.2249 \pm 0.0009$        & 1 & $-0.22$ & 0.16 & $-0.13$\\
    $A$ & $0.806  \pm 0.028$               & $-0.22$  & 1  & $-0.37$  & $-0.49$\\
    $ \overline{\rho}$ & $0.173\pm 0.016$  & 0.16 & $-0.37$ & 1 & 0.41  \\
    $ \overline{\eta}$  & $0.368\pm 0.024$ & $-0.13$ & $-0.49$ & 0.41 & 1 \\
\end{tabular}

\caption{Wolfenstein parameters obtained from a fit to
the CKM entries in Section \ref{sec:CKM_fit}.}
\label{tab:WolfResult}
\end{table}

\FloatBarrier

\subsection{Non-contaminated fit to the oblique parameters}
\label{sec:STU_fit}

For the determination of the oblique parameters a global fit of the EW observables is performed, removing the contribution from $R_b$. We use the same inputs as described in Ref.~\cite{deBlas:2022hdk}, but we remove $R_b$ and use the PDG value for $M_W$.

\begin{table}[h!]
\begin{center}
\begin{tabular}{c|c|c|c|c||c|c|c}
    \multirow{2}{*}{ } & \multirow{2}{*}{Value} & \multicolumn{3}{c||}{Correlation} & \multirow{2}{*}{Value} & \multicolumn{2}{c}{Correlation} \\\cline{3-5}\cline{7-8}
    & & $ S $ & $ T $ & $ U $ & & $S$ & $T$ \\ \hline
    $ S $ & 0.005 $\pm$ 0.096 & 1.00  &  0.91  &  -0.62 & 0.024 $\pm$ 0.076 & 1 & 0.91\\
     $ T $ & 0.042 $\pm$ 0.118 & 0.91  &  1.00  &  -0.84 & 0.075 $\pm$ 0.063 & 0.91 & 1\\
     $ U $ & 0.030 $\pm$ 0.091 & -0.62 &  -0.84 &  1.00 & & &\\
\end{tabular}
\end{center}
\caption{ Results for the fit of the oblique parameters $S$, $T$ and $U$, excluding the information from $R_b$.}
\label{tab:STU}
\end{table}

\FloatBarrier


\end{document}